\documentclass[journal]{IEEEtran}

\usepackage{amsmath, amssymb, amsthm}  
\allowdisplaybreaks

\usepackage{algorithm}      
\usepackage{algpseudocode}

\usepackage{graphicx}  
\usepackage{tabularx} 
\usepackage{caption}  
\usepackage{subcaption}  
\usepackage{booktabs}  
\usepackage{multirow}  
\usepackage{url}  
\usepackage{cite}  
\usepackage{color}  
\usepackage{xcolor}  
\usepackage{hyperref}  
\usepackage{siunitx}  
\usepackage{mathrsfs}  
\usepackage{mathtools}  
\usepackage{txfonts}  
\usepackage{fontawesome5}  
\usepackage{epstopdf}  
\usepackage{enumitem}  
\usepackage{pgfplots}  
\usepackage{verbatim}  
\usepackage{placeins}  
\usepackage{pdflscape}  
\usepackage{balance}  
\usepackage{lipsum}  
\usepackage{upgreek}  
\usepackage{amsfonts} 
\usepackage{bm}  

\makeatletter
\newenvironment{breakablealgorithm}
  {
   \begin{center}
     \refstepcounter{algorithm}
     \hrule height.8pt depth0pt \kern2pt
     \renewcommand{\caption}[2][\relax]{
       {\raggedright\textbf{\ALG@name~\thealgorithm} ##2\par}%
       \ifx\relax##1\relax 
         \addcontentsline{loa}{algorithm}{\protect\numberline{\thealgorithm}##2}%
       \else 
         \addcontentsline{loa}{algorithm}{\protect\numberline{\thealgorithm}##1}%
       \fi
       \kern2pt\hrule\kern2pt
     }
  }{
     \kern2pt\hrule\relax
   \end{center}
  }
\makeatother

\title{Trajectory Optimization for Cellular-Connected UAV in Complex Environment with Partial CKM}

\vspace{-2ex}

\author{
     \IEEEauthorblockN{Yuxuan Song, Haiquan Lu, Chiya Zhang,~\IEEEmembership{Member,~IEEE}, Beixiong Zheng, ~\IEEEmembership{Senior Member,~IEEE}, \\ and Yong~Zeng, ~\IEEEmembership{Fellow,~IEEE}}
\thanks{

Yuxuan Song and Yong Zeng are with the National Mobile Communications Research Laboratory, Southeast University, Nanjing 211189, China. Yong Zeng is also with the Purple Mountain Laboratories, Nanjing 211111, China (e-mail: \{220240906, yong\_zeng\}@seu.edu.cn). (Corresponding author: Yong Zeng.)

Haiquan Lu is with the School of Electronic and Optical Engineering, Nanjing University of Science and Technology, Nanjing 210094, China (e-mail: haiquanlu@njust.edu.cn).

Chiya Zhang is with the School of Electronic and Information Engineering, Harbin Institute of Technology, Shenzhen 518055, China (e-mail: zhangchiya@hit.edu.cn).

Beixiong Zheng is with the School of Microelectronics, South China University of Technology, Guangzhou, 511442, China (e-mail: bxzheng@scut.edu.cn).}}

\begin{document}
\maketitle

\vspace{-0.2in}

\begin{abstract}

Cellular-connected unmanned aerial vehicles (UAVs) are expected to play an increasingly important role in future wireless networks. To facilitate the reliable navigation for cellular-connected UAVs, channel knowledge map (CKM) is considered a promising approach capable of tackling the non-negligible co-channel interference resulting from the high line-of-sight (LoS) probability of air-ground (AG) channels. Nevertheless, due to measurement constraints and the aging of information, CKM is usually incomplete and needs to be regularly updated to capture the dynamic nature of complex environments. In this paper, we propose a novel trajectory design strategy in which UAV navigation and CKM completion are incorporated into a common framework, enabling mutual benefits for both tasks. Specifically, a cellular-connected UAV deployed in an urban environment measures the radio information during its flight and completes the CKM with Kriging interpolation. Based on the method of grid discretization and spherical approximation, a mixed-integer multi-objective optimization problem is formulated. The problem falls into the category of combinatorial mathematics and is essentially equivalent to determining an optimum sequence of grid points to traverse. Through proper mathematical manipulation, the problem is reformulated as variants of two classic models in graph theory, namely the shortest-path problem (SPP) and the traveling salesman problem (TSP). Two navigation strategies based on the two different models are proposed and thoroughly compared based on numerical results to provide implementable methods for engineering practice and reveal the trade-offs between UAV navigation and CKM completion. Simulation results reveal that the proposed navigation strategies can quickly expand the Pareto boundary of the problem and approach the performance of fully-known CKM.

\end{abstract}


\section{Introduction}

Commercial-grade unmanned aerial vehicles (UAVs) are expected to play an increasingly important role in future urban and rural environments. Thanks to their high mobility and deployment flexibility, UAVs can provide various types of service, including online video streaming, urban logistics, and agricultural plant protection, among others. Of all the available wireless technologies to support large-scale low-altitude UAV deployment, the cellular-connected approach is considered as the most effective way to realize beyond visual line-of-sight (LoS) UAV operations with fully expanded communication range \cite{zeng2018cellular}\cite{Lu2025}, thus opening new business opportunities and expanding human activities from 2-dimensional (2D) ground to 3-dimensional (3D) low-altitude aerial space.

\subsection{Related Works}

Despite the above-mentioned advantages, cellular communications for aerial users face new challenges. Due to the LoS-dominance characteristic of air-ground channels, aerial users generally possess better channel conditions with more ground base stations (BSs) compared to terrestrial users, causing non-negligible co-channel interference in both the uplink from the UAV to non-associated BSs and the downlink in the reverse direction. In order to address this issue, various interference mitigation techniques have been investigated.

Interference mitigation techniques in the uplink can be divided into signal processing-based methods\cite{LiuMulti} \cite{QFMei} and system optimization-based methods\cite{UpPower} \cite{D2DPro}. For the former, the authors in \cite{LiuMulti} proposed a spatial domain cancellation strategy based on digital beamforming to minimize the uplink interference from a single UAV to non-associated BSs serving ground users. For the latter, \cite{UpPower} and \cite{D2DPro} studied the uplink interference among multiple UAVs associated with different BSs and that within UAV swarms, which were respectively addressed via geographic programming (GP) and protocol-level optimization. In addition, non-orthogonal multiple access (NOMA)-based solutions have also found successful applications in this regard\cite{CoNOMA} \cite{CSINOMA} \cite{SGNOMA} \cite{NOMAPre}. Specifically, the authors in \cite{CoNOMA} proposed a cooperative NOMA scheme to maximize the weighted sum-rate of UAVs and ground users by exploiting existing backhaul links among BSs, leading to significant throughput gains compared to traditional orthogonal multiple access (OMA) and non-cooperative NOMA. To acquire accurate channel state information (CSI) required by NOMA-based interference cancellation, the authors in \cite{CSINOMA} proposed a two-stage semi-blind channel estimation algorithm exploiting the Doppler and modulation diversities, expanding the applicability of NOMA in the scenario where UAVs maneuver with high speed.

In terms of the interference management in the downlink, there has been a growing interest in employing local channel knowledge map (CKM) \cite{Zeng2021Mag}\cite{zeng2024tutorial} to facilitate the navigation of cellular-connected UAVs \cite{2019connectivity} \cite{2019Belgium} \cite{Aware} \cite{zhan2022energy} \cite{Zhang}. The basic idea of CKM is to learn the prior channel knowledge in 3D spatial or virtual locations such as path loss, delay, and angle-of-arrival (AoA). To facilitate the communication for cellular-connected UAVs, CKM can assist with the acquisition of finer-grained CSI\cite{Zeng2023ckm} and enable non-line-of-sight (NLoS) communication via link state identification \cite{2024Yang}, among others. The benefit of applying CKM for downlink interference management is twofold. First, the demand for the safe operation of cellular-connected UAVs puts a stringent requirement in terms of outage probability and communication latency on ground BSs that radiate control and non-payload communication (CNPC) signal in the downlink. As existing cellular networks are specially designed to serve terrestrial users, aerial communication coverage is still rather unreliable, especially in urban environments where signal propagation blockage and co-channel interference frequently occur due to complex physical and radio environments. Therefore, CKM is considered a favorable approach to navigating UAVs away from potential coverage holes through proper trajectory optimization and guiding the construction of aerial wireless networks in the future. Second, the definition of CKM relies on the method of grid discretization where the physical 2D or 3D space is discretized into square or cubic grid points so that the continuous radio environment involving an infinite number of spatial variables can be approximated into a set of finite spatial locations. This course of action not only makes the concept of CKM processable by digital computers, but also converts the path-planning problem into a more tractable form. Specifically, the problem is transformed from searching a continuous curve in the 3D space, which also involves dealing with an infinite number of spatial variables, into determining a traversal sequence of a finite number of spatial grid points. The problem thus falls into the category of combinatorial mathematics in which numerous classical and elegant algorithms are available. To be exact, in order to minimize the flight completion time while restricting the total duration of communication outage, the path-planning problem can be formulated as the shortest-path problem (SPP) in the field of graph theory and can be efficiently solved via the Dijkstra algorithm \cite{Zhang} and its derivative algorithms \cite{2019connectivity} \cite{2019Belgium} \cite{Aware}\cite{zhan2022energy}.

Meanwhile, due to various practical constraints, it is usually the case that the available CKM is only partially known, necessitating the process of reconstruction for further applications. In fact, there has been serious discussion regarding the minimum amount of data needed for the successful construction of a CKM \cite{xu2024much}. Thanks to the high mobility and operational flexibility of autonomous UAVs, it has become a compelling practice to deploy specialized cooperative UAVs to measure the electromagnetic information in the aerial space of interest\cite{3DTensor} \cite{AutoUAV} \cite{Semantics} \cite{measureCKM} \cite{hu20233d} and facilitate the construction of local 3D CKM, correspondingly empowering various task-driven applications such as the navigation of cellular-connected UAVs. However, deploying specialized UAVs to measure the partial CKM such as the practice in \cite{measureCKM} faces two challenges. First, given the broadness of the 3D low-altitude aerial space, it is rather challenging to design the trajectories of the measurement UAVs to achieve optimum information gain and even more time-consuming to actually implement the measurement process. Second, radio maps tend to be highly malleable by the subtle physical changes in the real world, requiring frequent deployments of measurement UAVs to timely update the CKM grid points with large age of information (AoI).

Due to the above reasons, it becomes more natural to integrate the construction of CKM and UAV navigation into a common framework, legitimizing the idea of allowing cellular-connected UAVs themselves to assume the responsibility to complete the partial, dynamically changing CKM and facilitate their own flights in the future. If they were allowed to deviate from their original flight tasks, cellular-connected UAVs could effectively update the local CKM and utilize the refined channel knowledge in a swift manner. In fact, several works\cite{zeng2021simultaneous} \cite{Federated} \cite{Fedarated3D} \cite{EnergyOnline} have already experimented with such an idea. In particular, a novel simultaneous navigation and radio mapping (SNARM) framework was proposed in \cite{zeng2021simultaneous} based on deep reinforcement learning which allows for the fast construction of the channel gain map (CGM) in urban air space.

The rationale above implies the formulation of a multi-objective optimization problem that integrates UAV navigation and CKM completion. The problem aims to strike a balance between these two components by exploring the trade-offs between them through proper performance metrics.

\subsection{Main Contributions}
\begin{enumerate}[label={\textbullet}]
\item Unlike previous works \cite{2019connectivity} \cite{zhan2022energy} \cite{Zhang} postulating that the CKM is fully known prior to the UAV's flights, the CKM in this paper is assumed to be partially known and needs to be completed based on the data harvested by cellular-connected UAVs along with Kriging interpolation. In order to combine UAV navigation and CKM completion into a common framework, a multi-objective optimization problem is formulated which aims to minimize flight completion time and outage duration, while maximizing the measurement information gain. Through proper mathematical manipulation, the problem can be reformulated as a graph optimization problem.
\item In contrast to previous works \cite{2019connectivity} \cite{2019Belgium} \cite{Aware} that employed the Manhattan distance or its derivatives to constitute the UAV's action space, in this paper, we use a spherical approximation method, which incorporates the Manhattan distance as a special case, to bestow a higher degree-of-freedom (DoF) on the UAV's trajectory and facilitate the weight mapping process in the proposed graph-based solutions.
\item In order to solve the graph-based navigation problem and complete the partial CKM efficiently, a sub-optimization problem based on the summed variances of estimation errors is formulated to determine the grid points having the highest measurement value, which constitute the waypoints of the UAV's trajectory. Two navigation strategies based on the SPP and the traveling salesman problem (TSP) in graph theory are investigated and compared to reveal the inner trade-offs among the three optimization objectives.
\end{enumerate}

The remainder of this paper is organized as follows. Section II introduces the system model with an emphasis on the definition of CKM and the notation of the UAV's waypoints. Section III discusses the spherical approximation method followed by the formulation of the multi-objective problem in its general form. In Section IV, two graph-based path planning strategies based on two different models are discussed. Section V compares the results based on the two strategies and reveals the trade-offs among the optimization objectives. Section VI concludes the paper.

\textit{Notations}: $\boldsymbol{A}^T$ denotes the transpose of matrix $\boldsymbol{A}$; $\|\boldsymbol{x}\|$ denotes the Euclidean norm of vector $\boldsymbol{x}$; $|S|$ refers to the cardinality of set $S$; $\lceil x \rceil$ is the ceiling function for $x$; $\mathcal{X} \setminus \mathcal{Y}$ denotes the set of elements that are in set $\mathcal{X}$ but not in set $\mathcal{Y}$; $\preceq$ is the componentwise inequality; $\mathbb{R}$ denotes the set of real numbers, and $\mathbb{N}$ denotes the set of natural numbers.

\section{System Model}

As illustrated in Fig. 1, the considered system consists of one cellular-connected UAV operating in an urban environment with high-rise buildings and $M\geq1$ ground BSs available for association with the UAV during its flight. Define $\mathcal {M}=\{1,2,...M\}$. The ground BSs' coordinates are denoted as ${B_m = [a_m, b_m, H_m]^T \in \mathbb{R}^{3 \times 1}, m\in\mathcal{M}}$. The mission of the UAV is to fly from a starting point $u_s = [x_s, y_s, z_s]^T \in \mathbb{R}^{3 \times 1}$ to an end point $u_e = [x_e, y_e, z_e]^T \in \mathbb{R}^{3 \times 1}$ while maintaining reliable communication with ground cellular networks with the help of the CKM. The CKM is assumed to be only partially known a priori, and therefore the UAV needs to measure as much unknown channel knowledge as possible during its flight to complete the partial CKM.

\begin{figure}[h!]
    \centering
    \includegraphics[width=9cm,height=5cm]{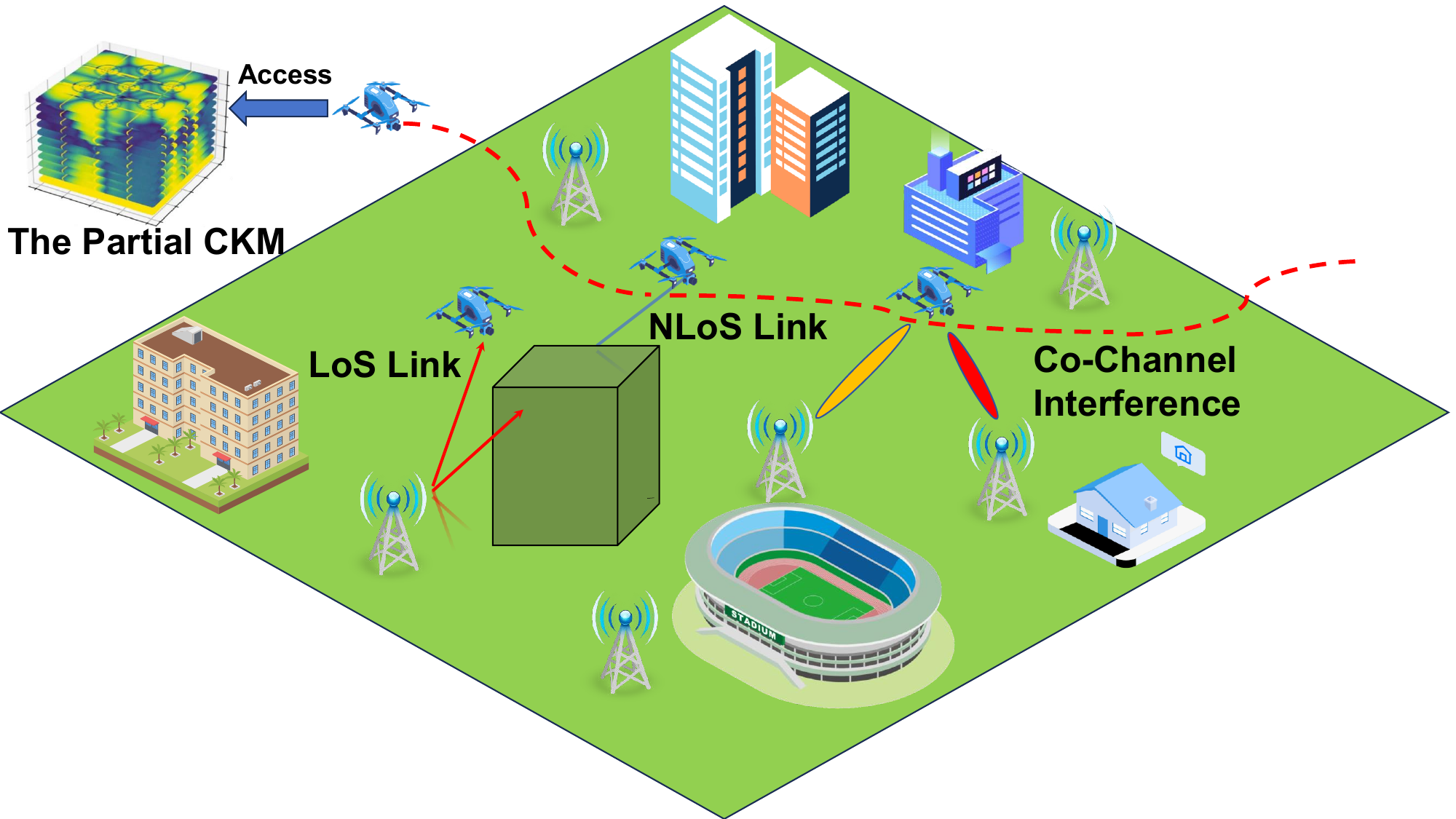}
    \caption{Cellular-connected UAV with partial CKM.}
\end{figure}

The UAV is assumed to operate in a 3D aerial space which is denoted as $[x_L, x_U] \times [y_L, y_U] \times [z_L, z_U]$. Since this paper focuses on communication-related optimization objectives, it is assumed that the aforementioned 3D aerial space is clear of physical obstacles, and therefore every location in the 3D space is reachable. Without loss of generality, the definition of CKM and the UAV's trajectory start from the continuous domain. Specifically, the UAV's continuous trajectory is mathematically characterized as ${\boldsymbol{u} (t)=[x(t), y(t), z(t)]^T \in \mathbb{R}^{3 \times 1},t\in[0,T]}$. The instantaneous received signal strength (RSS) from BS $m$ to the UAV is denoted as $P_m(\boldsymbol{u}(t)),m\in\mathcal{M}$, which is explicitly characterized as\cite{zeng2021simultaneous}
\begin{equation} 
P_m(\boldsymbol{u}(t)) = \overline{P}_m B_m(\boldsymbol{u}(t)) G_m(\boldsymbol{u}(t)) h_m(t),
\end{equation} where $\overline{P}_m$ denotes the average transmit power of BS $m$, $B_m(\boldsymbol{u}(t))$ and $G_m(\boldsymbol{u}(t))$ respectively account for the large-scale channel gain and the beampattern gain from BS $m$ to location $\boldsymbol{u}(t)$. $h_m(t)$ is the small-scale fading factor with normalized average power, i.e., $\mathbb{E}[h_m(t)]=1$.

Define an indicator function $b(\boldsymbol{u}(t))$. $b(\boldsymbol{u}(t))=m, m\in\mathcal{M}$ means that the UAV is associated with BS $m$ at time instant $t$. Thus, the received signal-to-interference-plus-noise ratio (SINR) is given by
\begin{equation}
\gamma(\boldsymbol{u}(t)) = \frac{P_{b(\boldsymbol{u}(t))}(\boldsymbol{u}(t))}{\sum_{m^{\prime} \in \mathcal{M} \backslash \{b(\boldsymbol{u}(t))\}} P_{m'}(\boldsymbol{u}(t)) + \sigma_n^2},\\
\end{equation} where $\sigma_n^2$ denotes the noise power at the receiver. In reality, the instantaneous value of $\gamma(\boldsymbol{u}(t))$ changes rapidly on the time scale of the channel coherence time. In order to store and access the CKM in an offline manner, the CKM \cite{zhan2022energy}is defined as the expected SINR at each spatial location, given by
\begin{equation}
\mathbb{E}[\gamma(\boldsymbol{u}(t))] \overset{(a)}{\geq}
    \frac{\overline{P}_{b(\boldsymbol{u}(t))}\cdot B_{b(\boldsymbol{u}(t))}(\boldsymbol{u}(t))\cdot G_{b(\boldsymbol{u}(t))}(\boldsymbol{u}(t))}
{\sum_{m^{\prime} \in \mathcal{M} \backslash \{{b(\boldsymbol{u}(t))}\}} \overline{P}_{m^{\prime}} \cdot B_{m^{\prime}}(\boldsymbol{u}(t))\cdot G_{m^{\prime}}(\boldsymbol{u}(t)) + \sigma_n^2}, \label{SINR}
\end{equation} where (a) follows from the Jenson's inequality and the fact that $\frac{1}{x}$ is convex for $x>0$. In addition, the associated BS in each time instant is selected as the BS that can yield the maximum expected SINR, i.e.,
\begin{equation}
b(\boldsymbol{u}(t)) = \arg\max_{m \in \mathcal{M}} \frac{\overline{P}_m\cdot B_m(\boldsymbol{u}(t))\cdot G_m(\boldsymbol{u}(t))}
{\sum_{m^{\prime} \in \mathcal{M} \backslash \{m\}} \overline{P}_{m^{\prime}} \cdot B_{m^{\prime}}(\boldsymbol{u}(t))\cdot G_{m^{\prime}}(\boldsymbol{u}(t)) + \sigma_n^2}. \\
\end{equation}

In order to digitally store the CKM and the UAV's waypoints, the 3D aerial space is discretized into cubic grid points. Specifically, all three spatial dimensions are uniformly discretized with granularity $\Delta_D$, and the 3D space is approximated to $I \cdot J \cdot K$ grid points, where $I = \lceil \frac{x_U-x_L}{\Delta_{D}} \rceil$, $J = \lceil \frac{y_U-y_L}{\Delta_{D}} \rceil$, and $K = \lceil \frac{z_U-z_L}{\Delta_{D}} \rceil$. Three alphabets are subsequently defined as $\mathcal{X} = \{1, 2, \dots, I\}$, $\mathcal{Y} = \{1, 2, \dots, J\}$, $\mathcal{Z} = \{1, 2, \dots, K\}$. Therefore, each considered grid point can be mathematically indexed as ${u_{i,j,k}^D}, i \in \mathcal {X}, j \in \mathcal {Y},k \in \mathcal {Z}$. Note that $u_{i,j,k}^D$ is a logical variable representing the logical existence of a specific grid point $u_{i,j,k}^D$. Define a tensor field $\boldsymbol{u}^D:\mathbb{N}^3\rightarrow\mathbb{R}^{3\times 1}$, the Cartesian coordinate of the grid point $u_{i,j,k}^D$'s geographic center, stored as a $3 \times 1$ vector, can be denoted as 

\begin{equation}
\boldsymbol{u}^D(i,j,k)=\Delta_D{[(i-\frac{1}{2}),(j-\frac{1}{2}) ,(k-\frac{1}{2})]}^T,i \in \mathcal {X}, j \in \mathcal {Y},k \in \mathcal {Z}. 
\end{equation} When the grid granularity is sufficiently small, the expected SINR within a grid point can be approximately regarded as constant, i.e.,
\begin{equation}
\gamma(\boldsymbol{u}(t)) = \gamma(\boldsymbol{u}^D(i,j,k)),
\end{equation} if
$ |\boldsymbol{u}(t) - \boldsymbol{u}^D(i, j, k)|  \preceq \frac{\Delta_D}{2} [1,1,1]^T$. When $\mathbb{E}[\gamma(\boldsymbol{u}^D(i,j,k))] <\gamma_{th}$, grid point ${u_{i,j,k}^D}$ is considered to be in outage.

Previous works that assumed full knowledge of CKM\cite{2019connectivity} \cite{2019Belgium} \cite{Aware} \cite{zhan2022energy} \cite{Zhang} only have to deal with optimization objectives that are on the time scale of a single round of flight, such as completion time and outage duration. Nevertheless, for the additional task of CKM completion considered in this paper, the number of grid points measured in a single round is quite limited, suggesting that the effect of CKM completion must be discussed on a longer time scale. Therefore, the UAV is assumed to perform $R$ rounds of flight to iteratively update the global CKM. The UAV's trajectory in each round of flight consists of $N$ waypoints, or equivalently $N-1$ connected line segments. The UAV's $n^{th}$ waypoint in the $r^{th}$ round of flight is denoted as $\boldsymbol{u}_r[n]=\boldsymbol{u}^D(i_n^r,j_n^r,k_n^r)$ where $i_n^r \in \mathcal {X}, j_n^r \in \mathcal {Y},k_n^r \in \mathcal {Z},1 \leq n \leq N, 1 \leq r \leq R$. The $n^{th}$ line segment in the $r^{th}$ round of flight can be denoted in the parametric form as 

\begin{equation}
\begin{aligned}
\boldsymbol{u}_r[n+1]-\boldsymbol{u}_r[n]=
\begin{cases}
x = (i_n^r-\frac{1}{2})\Delta_D + t(i_{n+1}^r - i_n^r)\Delta_D \\
y = (j_n^r-\frac{1}{2})\Delta_D + t(j_{n+1}^r - j_n^r)\Delta_D \\
z = (k_n^r-\frac{1}{2})\Delta_D + t(k_{n+1}^r - k_n^r)\Delta_D
\end{cases}  
\\ \text{where } 0 \leq t \leq 1, 1 \leq n \leq N-1,\text{and }1 \leq r \leq R. 
\end{aligned}
\end{equation}

As defined in \eqref{SINR}, the CKM in this paper is mainly determined by the large-scale propagation loss $B_m$ and the beampattern gain $G_m$. Therefore, the ground-truth CKM varies with environmental changes such as the construction of new buildings and antenna array calibration, suggesting that the completion of CKM is an ever-lasting process requiring constant information update. In this paper, the UAV's trajectory is designed in an offline manner, and the ground-truth CKM is assumed to be invariant during every $R$ rounds of flight. After every $R$ rounds, the partial CKM is updated globally, and the unmeasured grid points in the new partial CKM can be selected based on the age of information (AoI) and its variant metrics\cite{AoIUpdate}\cite{AoI}\cite{CKMUpdate}. The more challenging online optimization incorporating real-time channel variation will be considered in our future work.

\section{Problem Formulation}

In this paper, we consider three short-term optimization objectives, namely the completion time, the outage duration, and the number of measured grid points in a single round flight, and one long-term objective of minimizing the global MSE after $R$ rounds of flight. The completion time and the outage duration are equivalently expressed as the total traveling distance and the total length of intersection between the UAV's trajectory and grid points in outage. In order to characterize the total outage duration, a spherical approximation method is adopted. Take the 2D plane as a special case. As illustrated in Fig. 2, each grid point is approximated by a circle with a radius of $R_D=\Delta_D/2$. The distance from the center of the grid point ${u_{i,j,k}^D}$, which is $\boldsymbol{u}^D(i,j,k)$, to the $n^{th}$ line segment in the $r^{th}$ round $\boldsymbol{u}_r[n+1]-\boldsymbol{u}_r[n]$ is $dist^{n,r}_{i,j,k}=\frac {\| (\boldsymbol{u}_r[n]-\boldsymbol{u}^D(i,j,k))\times (\boldsymbol{u}_r[n+1]-\boldsymbol{u}_r[n])\|}{\|\boldsymbol{u}_r[n+1]-\boldsymbol{u}_r[n]\|}$. Therefore, the approximated length of intersection between the UAV's trajectory and the grid point ${u_{i,j,k}^D}$ that is in outage is $L^{n,r}_{i,j,k}=2\sqrt{R_D^2-{(dist^{n,r}_{i,j,k})}^2}$. Note that if the restriction $\|\boldsymbol{u}_r[n+1]-\boldsymbol{u}_r[n]\|=\Delta_D$ is posed to the UAV's trajectory, the distance between any $\boldsymbol{u}_r[i], \boldsymbol{u}_r[j],i\neq j$ is the standard Manhattan distance, which is a special case of the proposed approximation method.

\begin{figure}[h!]
    \centering
    \includegraphics[width=9cm,height=5cm]{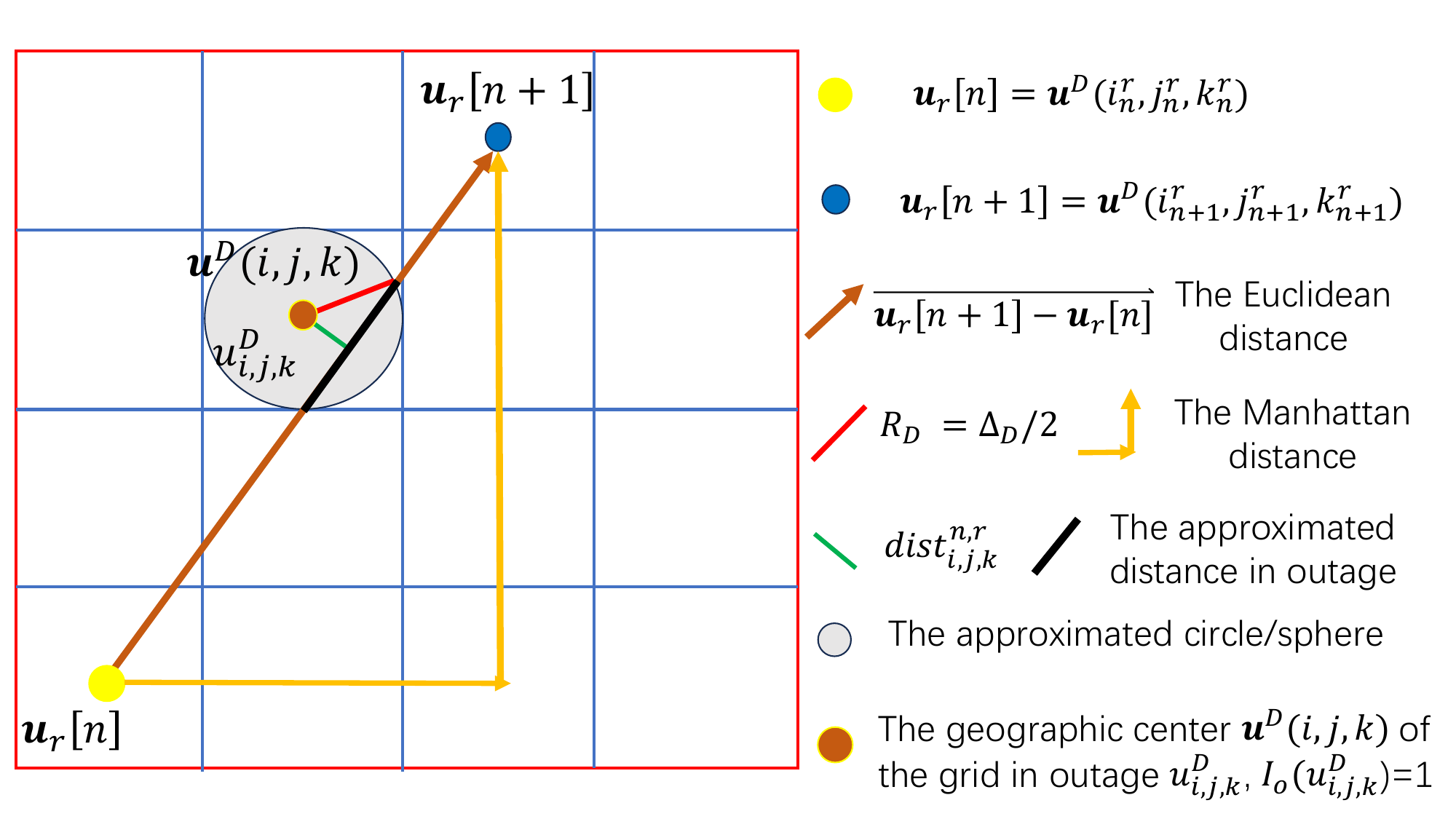}
    \caption{An illustration of the spherical approximation.}
\end{figure}

In order to mathematically characterize the optimization objectives, three indicator functions are defined as follows.

\begin{align}
I_T(dist^{n,r}_{i,j,k}) &= 
\begin{cases}
1, &\text{if} \text{\quad $dist^{n,r}_{i,j,k}<R_D$}  \\
0.  &
\end{cases} \\   
I_O(u_{i,j,k}^D) &= 
\begin{cases}
1, & \text{if} \quad \gamma(\boldsymbol{u}^D(i,j,k)) <\gamma_{th}\\
0. & 
\end{cases} \\
I_M^r(u_{i,j,k}^D) &= 
\begin{cases}
1, & \text{if grid } u_{i,j,k}^D \text{has not been measured} \\
&\text{prior to the $r^{th}$ round of flight}\\
0. &
\end{cases}
\end{align}

The first indicator $I_T(dist^{n,r}_{i,j,k})$ is defined to indicate if the spatial grid point $u_{i,j,k}^D$ intersects the $n^{th}$ line segment in the $r^{th}$ round. When it does, $I_T(dist^{n,r}_{i,j,k})=1$. The total $(N-1)\cdot I\cdot J\cdot K$ indicators jointly characterize the grid points traversed by the UAV in the $r^{th}$ round. In terms of the second indicator $I_O(u_{i,j,k}^D)$, it indicates the communication quality within the grid point $u_{i,j,k}^D$. Note that the communication quality of a particular grid point $u_{i,j,k}^D$ has only to do with the environmental setting and is independent of the UAV's trajectory, and therefore no superscript is added to this indicator. As for the last indicator $I_M^r(u_{i,j,k}^D)$ \cite{AoI}, it specifies the measurement status of each grid point. If the grid point $u_{i,j,k}^D$ has never been measured before the $r^{th}$ round of flight, the indicator equals 1 to encourage the UAV to measure it. Note that the measurement status of a grid point is independent of the communication quality within it, meaning that an unmeasured grid point can be either of high communication quality or in communication outage. The SINR in unmeasured grid points is estimated based on Kriging interpolation, and an unmeasured grid point is considered to be in outage if its estimated SINR is below the designated outage threshold.

As pointed out in Section II, the completion of CKM is an on-going process which requires the continuous update of measurement information from consecutive rounds of flight, with the minimum sacrifice of the completion time and outage duration. Therefore, on the time scale of a single round of flight, the optimization objectives are formulated as follows.

\begin{align}
T_r &=\sum_{n = 1}^{N-1}\|\boldsymbol {u}_r[n+1]-\boldsymbol {u}_r[n]\|,\\
O_r &= \sum_{n = 1}^{N-1}\sum_{i = 1}^{I} \sum_{j = 1}^{J} \sum_{k = 1}^{K} I_T(dist^{n,r}_{i,j,k}) \cdot I_O(u_{i,j,k}^D)\cdot L^{n,r}_{i,j,k}, \label{outage} \\ 
M_r &= \sum_{n = 1}^{N-1}\sum_{i = 1}^{I} \sum_{j = 1}^{J} \sum_{k = 1}^{K} I_T(dist^{n,r}_{i,j,k}) \cdot I_M^{r}(u_{i,j,k}^D).
\end{align}

$T_r$ stands for the total length of the UAV's trajectory in the $r^{th}$ round of flight, which corresponds to the completion time. $O_r$ refers to the total length of the UAV's trajectory in the $r^{th}$ round of flight that intersects grid points in outage, corresponding to the total outage duration. The last objective $M_r$ characterizes the total number of unmeasured grid points traversed by the UAV during its $r^{th}$ round of flight.

For the long-term objective, the UAV aims to minimize the mean-square error (MSE) of the global CKM based on Kriging interpolation. Denote the estimated SINR in grid point $u_{i,j,k}^D$ as $\hat{\gamma}(\boldsymbol{u}^D(i,j,k))$, the MSE of the global CKM after all $R$ rounds of flight is 

\begin{equation}
\text{MSE}= \sum_{i = 1}^{I} \sum_{j = 1}^{J} \sum_{k = 1}^{K} \frac{1}{I\cdot J\cdot K} ( {\gamma}(\boldsymbol{u}^D(i,j,k))-\hat{\gamma}(\boldsymbol{u}^D(i,j,k)))^2.
\end{equation}

With the definitions above, the multi-objective optimization problem can be formulated as a mixed-integer programming problem as follows.

\begin{align}
\textup{(P1)}\quad
\min_{\boldsymbol {u}_r[n], 1\leq n\leq N,1\leq r\leq R} 
&\quad T_r,O_r,\text{MSE}\\
\max_{\boldsymbol {u}_r[n], 1\leq n\leq N,1\leq r\leq R} &\quad M_r\\
\text{s.t.}  \quad 
&\quad \boldsymbol {u}_r[n] = \boldsymbol{u}^D(i_n^r,j_n^r,k_n^r), \notag \\
&\quad \exists i_n^r \in \mathcal {X}, j_n^r \in \mathcal {Y},k_n^r \in \mathcal {Z}. 
\end{align}

Note that no additional constraints are needed to reflect the straight-line flight assumption because it has been fully characterized in $T_r$ and $O_r$. The problem is equivalent to determining a discrete sequence of spatial locations, and thus falls into the category of combinatorial mathematics. In addition, although the ultimate objective of CKM completion is to minimize the global MSE based on the measurement data harvested from all $R$ rounds of flight, the design of the UAV's trajectory must be performed on a single-flight basis. Therefore, the introduction of the objective $M_r$ to be maximized in each round is well justified.

\section{Proposed Algorithms}

In this section, we propose two strategies to tackle problem (P1) based on the SPP and TSP in graph theory. Table 1 summarizes the key symbol definitions in graph theory.

\begin{table}[htbp]
\centering
\caption{Summary of the key symbol definitions}
  \newcolumntype{L}{>{\bfseries\centering\arraybackslash}p{0.29\linewidth}}
\begin{tabularx}{\linewidth}{ 
    L  
    X  
  }
\hline
\textbf{Symbol} & \textbf{Definition} \\ \hline
$V(G)$ & The nonempty vertex set of undirected graph $G$.    \\ \hline
$V(D)$ & The nonempty vertex set of directed graph $D$.    \\ \hline
$E(G)$ & The edge set of undirected graph $G$.  \\ \hline
$E(D)$ & The edge set of directed graph $D$.  \\ \hline
$e=(v_i,v_j)$ & The edge connecting vertices $v_i, v_j\in V$. For directed graphs, $(v_i,v_j)\neq(v_j,v_i)$.   \\ \hline
$\boldsymbol{w}=\sum_{i=1}^{|E|}y_i\boldsymbol{a}_i$ & The weight mapping function characterized by a bases of $|E|$ orthogonal functions $\{\boldsymbol{a}_1,\boldsymbol{a}_2,...\boldsymbol{a}_{|E|}\}$, where $\quad \boldsymbol{a}_i(e_j)=
\begin{cases}
1, \quad i=j\\
\\
0, \quad i\neq j
\end{cases}$, with the weight on each $e_i$ being $y_i$.
\\ \hline
$G=(V(G), E(G),\boldsymbol{w})$ & The full definition of undirected weighted graph $G$.  \\ \hline
$D=(V(D), E(D),\boldsymbol{w})$ & The full definition of directed weighted graph $D$.  \\ \hline
$d_G(v)$ & The vertex degree of vertex $v\in V(G)$. \\ \hline
$d^+_D(v)$ &The out-degree of vertex $v\in V(D)$. \\ \hline
$d^-_D(v)$ &The in-degree of vertex $v\in V(D)$. \\ \hline
$K_N(\mathcal{S})$ & Undirected complete graph with vertex set $\mathcal{S}$, $|\mathcal{S}|=N$.\\ \hline
\end{tabularx}
\end{table}

\subsection{Kriging Interpolation-based CKM Completion}

The process of measuring previously unmeasured grid points and incorporating them into the vertex set for navigation expands the Pareto boundary of the proposed multi-objective problem. After revealing the communication status in previously unknown areas, the set of grid points beneficial to minimizing the completion time and outage duration can be enlarged, and a higher DoF for trajectory design can be obtained. In this paper, the ground-truth CKM is assumed to be invariant during the $R$ rounds of flight. Therefore, the static Kriging focusing on spatial correlation\cite{Semantics} \cite{measureCKM}is employed to estimate the SINR in unmeasured grid points. Future works considering dynamic environmental changes could use Kalman-filter-based\cite{DynamicKF}\cite{DistributedKriging} strategies to actively track the change in channel information.

The fundamental principle of Kriging is to exploit the spatial correlation between measured grid points and unmeasured ones, so as to find the unbiased minimum-variance estimator for the SINR in unmeasured grid points. With a slight abuse of notations, both the index for a grid point $\boldsymbol{u}_{i_0,j_0,k_0}^{D}$ and its geometric center $\boldsymbol{u}^D(i_0,j_0,k_0)$ are denoted as $\boldsymbol{u}_0$. Denote the estimated SINR in $\boldsymbol{u}_0$ as $\hat{\gamma}(\boldsymbol{u}_0)$. The Kriging interpolation for $\hat{\gamma}(\boldsymbol{u}_0)$ is 

\begin{equation}
\hat{\gamma}(\boldsymbol{u}_0) = \sum_{i=1}^{N} \lambda_i {\gamma}(\boldsymbol{u}_i), \label{Kriging Formula}
\end{equation} which is a linear combination of the SINR in $N$ measured grid points $\boldsymbol{u}_1,\boldsymbol{u}_2,...\boldsymbol{u}_N$. To apply Kriging, the CKM is considered as a spatial random process where the SINR value ${\gamma}(\boldsymbol{u}_i)$ in each grid point $\boldsymbol{u}_i$ is modeled as a random variable, and the subsequent mathematical deduction is based on the assumption of the first- and second-order stationarity. Specifically, the first-order stationarity requires that all the mean SINR value in each grid point be identical, i.e., $\mathbb{E}[{\gamma}(\boldsymbol{u}_i)] = \mu, \quad i \in \{ 0, 1, \ldots, IJK\}$. The second-order stationariy demands that the covariance of two spatial random variables ${\gamma}(\boldsymbol{u}_i)$ and ${\gamma}(\boldsymbol{u}_j)$ depend only on their Euclidean distance $d_{i,j}$. Each spatial random variable has the same variance, i.e., $\text{Var}\left[ {\gamma}(\boldsymbol{u}_i) \right] = \sigma_\gamma^2, \quad i \in \{ 0, 1, \ldots, IJK\}$. Define $R_i={\gamma}(\boldsymbol{u}_i)-\mu$, then $\mathbb E\left[ R_i \right] = 0, \text{Var}\left[ R_i \right] = \sigma^2_\gamma$.

Denote the SINR estimation error in $\boldsymbol{u}_0$ as $ \epsilon_{\boldsymbol{u}_0}=\hat{\gamma}(\boldsymbol{u}_0) - {\gamma}(\boldsymbol{u}_o)$. In order for the estimator to be unbiased, the mean value of $\hat{\gamma}(\boldsymbol{u}_0)$ should be equal to ${\gamma}(\boldsymbol{u}_o)$, i.e., $\mathbb E[\epsilon_{\boldsymbol{u}_0}] =0$. Given \eqref{Kriging Formula}, we have
\begin{equation}
\mathbb E\left( \sum_{i=1}^{N} \lambda_i {\gamma}(\boldsymbol{u}_i) - {\gamma}(\boldsymbol{u}_0) \right) = 0.
\end{equation}

The first-order stationarity condition is then utilized to get
\begin{equation}
\sum_{i=1}^{N} \lambda_i = 1 \label{sum=1}.   
\end{equation}

Define the variance of $ \epsilon_{\boldsymbol{u}_0}$ as $\sigma_{\boldsymbol{u}_0}^2$. Then, we have

\begin{equation}
\begin{split}
\sigma_{\boldsymbol{u}_0}^2 &= \text{Var}\left( \sum_{i=1}^{N} \lambda_i {\gamma}(\boldsymbol{u}_i) - {\gamma}(\boldsymbol{u}_0) \right) \\
&= \text{Var}\left( \sum_{i=1}^{N} \lambda_i {\gamma}(\boldsymbol{u}_i) \right) - 2 \text{Cov}\left( \sum_{i=1}^{n} \lambda_i {\gamma}(\boldsymbol{u}_i), {\gamma}(\boldsymbol{u}_0) \right) +\text{Var}({\gamma}(\boldsymbol{u}_0))  \\
&= \text{Var}\left( \sum_{i=1}^{N} \lambda_i {\gamma}(\boldsymbol{u}_i) \right) - 2 \text{Cov}\left( \sum_{i=1}^{n} \lambda_i {\gamma}(\boldsymbol{u}_i), {\gamma}(\boldsymbol{u}_0) \right) 
\\  &\quad +\text{Cov}\left({\gamma}(\boldsymbol{u}_0), {\gamma}(\boldsymbol{u}_0)\right)  \\
&= \sum_{i=1}^{N} \sum_{j=1}^{N} \lambda_i \lambda_j \text{Cov}({\gamma}(\boldsymbol{u}_i), {\gamma}(\boldsymbol{u}_j)) - 2 \sum_{i=1}^{N} \lambda_i \text{Cov}({\gamma}(\boldsymbol{u}_i), {\gamma}(\boldsymbol{u}_0)) 
\\  &\quad+ \text{Cov}({\gamma}(\boldsymbol{u}_0), {\gamma}(\boldsymbol{u}_0)). \label{1}
\end{split}
\end{equation}

Let $C_{i,j}=\text{Cov}({\gamma}(\boldsymbol{u}_i), {\gamma}(\boldsymbol{u}_j))=\text{Cov}(R_i, R_j)$. Then \eqref{1} can be simplified as 

\begin{equation}
\sigma_{\boldsymbol{u}_0}^2 = \sum_{i=1}^{N} \sum_{j=1}^{N} \lambda_i \lambda_j C_{ij} - 2 \sum_{i=1}^{N} \lambda_i C_{i0} + C_{00}.    
\end{equation}

The spatial correlation is quantized via the semivariogram, which is defined as $r_{ij}=\sigma_\gamma^2-C_{ij}$, and we have 

\begin{equation}
\begin{split}
\sigma_{\boldsymbol{u}_0}^2 &= \sum_{i=1}^{N} \sum_{j=1}^{N} \lambda_i \lambda_j (\sigma_\gamma^2 - r_{ij}) - 2 \sum_{i=1}^{N} \lambda_i (\sigma^2 - r_{i0}) 
\\  &\quad + \sigma_\gamma^2 - r_{00} \\
&= \sum_{i=1}^{N} \sum_{j=1}^{N} \lambda_i \lambda_j \sigma_\gamma^2 - \sum_{i=1}^{N} \sum_{j=1}^{N} \lambda_i \lambda_j r_{ij} - 2 \sum_{i=1}^{N} \lambda_i \sigma_\gamma^2 
\\  &\quad + 2 \sum_{i=1}^{N} \lambda_i r_{i0} + \sigma_\gamma^2 - r_{00}. \label{2}
\end{split}
\end{equation}

With \eqref{sum=1}, \eqref{2} is reduced to 

\begin{equation}
\begin{split}
\sigma_{\boldsymbol{u}_0}^2 &= \sigma_\gamma^2 - \sum_{i=1}^{N} \sum_{j=1}^{N} \lambda_i \lambda_j r_{ij} - 2\sigma_\gamma^2 + 2 \sum_{i=1}^{N} \lambda_i r_{i0} 
\\  &\quad+ \sigma_\gamma^2 - r_{00} \\
&= 2 \sum_{i=1}^{N} \lambda_i r_{i0} - \sum_{i=1}^{N} \sum_{j=1}^{N} \lambda_i \lambda_j r_{ij} - r_{00}.
\end{split}
\end{equation}

In order to find the optimum set of $\{\lambda_i\}$ that minimizes $\sigma_{\boldsymbol{u}_0}^2$ subject to \eqref{sum=1}, the lagrangian function is formulated as

\begin{equation}
\sigma_{\boldsymbol{u}_0}^2 + 2\upsilon \left( \sum_{i=1}^{N} \lambda_i - 1 \right),   \label{lagrangian} 
\end{equation} where $\upsilon$ denotes the lagrangian multiplier for the equality constraint \eqref{sum=1}. By setting the derivatives of \eqref{lagrangian} with respect to each $\lambda_i$ and the multiplier $\upsilon$ to zeros, we have

\begin{equation}\label{Matrix}
\underbrace{
\begin{bmatrix}
r_{11} & r_{12} & \cdots & r_{1N} & 1 \\
r_{21} & r_{22} & \cdots & r_{2N} & 1 \\
\vdots & \vdots & \ddots & \vdots & \vdots \\
r_{N1} & r_{N2} & \cdots & r_{NN} & 1 \\
1 & 1 & \cdots & 1 & 0
\end{bmatrix}
}_{\boldsymbol{R}}
\underbrace{
\begin{bmatrix}
\lambda_1 \\
\lambda_2 \\
\vdots \\
\lambda_N \\
-\upsilon
\end{bmatrix}
}_{\boldsymbol{\Lambda}}
=
\underbrace{
\begin{bmatrix}
r_{1\mathbf{0}} \\
r_{2\mathbf{0}} \\
\vdots \\
r_{N\mathbf{0}} \\
1
\end{bmatrix}
}_{\boldsymbol{r}_\mathbf{0}}. 
\end{equation} 

The relationship between $r_{ij}$ and $d_{ij}$ can be quantized through the semivariogram function, which is often approximated via data fitting. In this paper, the exponential model is employed
\begin{equation}
r(d) = C_0 + C\left( 1 - e^{-d/a} \right), \label{semivariogram}
\end{equation} where $d$ is the distance between different grid points, and $C_0,C,a$ are constants fitted by measurement data.

\subsection{The SPP-based solution}

One straightforward strategy for addressing a multi-objective optimization problem is to properly weight all objectives and sum them up, leading to a single-objective optimization problem. For (P1), the ultimate optimization objective to be minimized in each round can be coarsely denoted as $T_r+\mu_1O_r+\mu_2M_r$, where $\mu_1$ and $\mu_2$ denote the non-negative and non-positive weights, respectively. With proper definition of a directed graph $D_s=(V(D_s),E(D_s),\boldsymbol{w}_s)$, where the subscript $s$ refers to ``SPP-based'', the problem can be reformulated as an SPP and efficiently solved through mature polynomial algorithms.

\begin{figure}[h!]
    \centering
    \includegraphics[width=9cm,height=5cm]{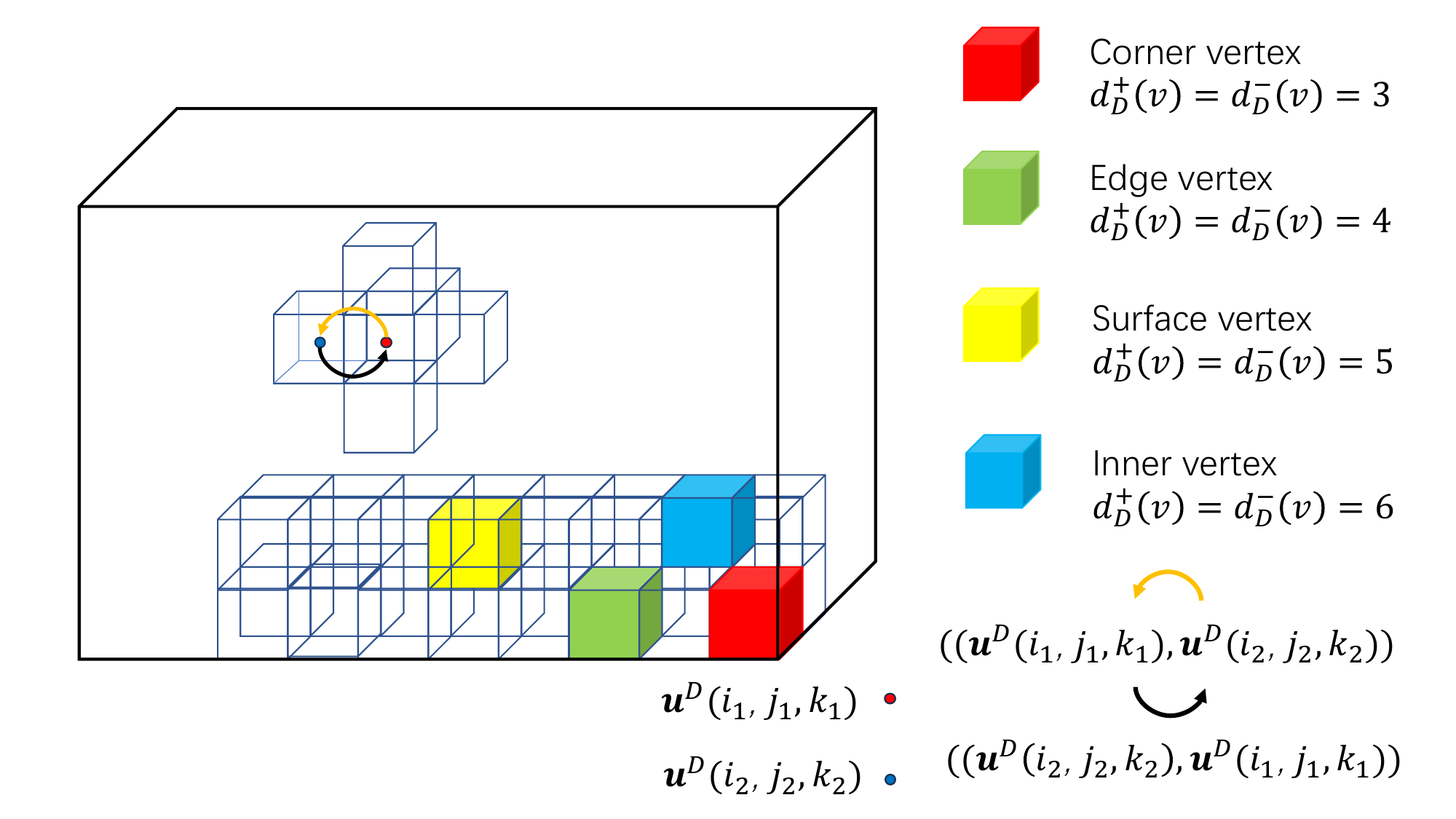}
    \caption{An illustration of $D_s$.}
\end{figure}

In order to mathematically formulate the SPP, the weighted objective function needs to be quantized based on the grid discretization introduced in Section II. Specifically, the 3D aerial space as a cuboid is replaced by $I\cdot J\cdot K$ cubic grid points. Each grid point can be considered as a vertex and they jointly constitute $V(D_s)$. Two spatial grid points $\boldsymbol{u}_{i_1,j_1,k_1}^D$ and $\boldsymbol{u}_{i_2,j_2,k_2}^D$ are adjacent to each other if and only if $\|\boldsymbol{u}^D(i_1,j_1,k_1)-\boldsymbol{u}^D(i_2,j_2,k_2))\|=\Delta_D$. The UAV can only fly to one of the grid points adjacent to its current grid, which restricts its action space to 3 to 6 directions, depending on the location of the current grid point as illustrated in Fig. 3. Under this setup, the trajectory of the UAV is a Manhattan trajectory, which is a special case of the spherical approximation, as pointed out in Section III. Each vertex forms two directed edges with each of its adjacent vertices. In terms of the weight on each edge, it should be the sum of three terms corresponding to the three objectives. Therefore, $D_s=(V(D_s),E(D_s),\boldsymbol{w}_s)$ can be expressed as

\begin{align}
&V(D_s)=\left\{\boldsymbol{u}^D(i,j,k) \mid i \in I,\ j \in J,\ k \in K \right\}  
\\&E(D_s)=\{(\boldsymbol{u}^D(i_a,j_a,k_a),\boldsymbol{u}^D(i_b,j_b,k_b))\mid  \notag
\\&\|\boldsymbol{u}^D(i_a,j_a,k_a)-\boldsymbol{u}^D(i_b,j_b,k_b)\|=\Delta_D\} 
\\ 
&\boldsymbol{w}_s(\boldsymbol{u}^D(i_a,j_a,k_a),\boldsymbol{u}^D(i_b,j_b,k_b))= w_{a,b}^s= \notag\\
&\underbrace{\Delta_D}_{T_r} \notag \\
& \underbrace{+\frac{1}{2}\mu_1\Delta_D I_O(\boldsymbol{u}_{i_a,j_a,k_a}^D) +\frac{1}{2}\mu_1\Delta_D I_O(\boldsymbol{u}_{i_b,j_b,k_b}^D)}_{O_r} \notag \\ 
&+\underbrace{\mu_2\Delta_DI_M(\boldsymbol{u}_{i_b,j_b,k_b}^D)}_{M_r},\quad \mu_1 \geq 0 \quad \mu_2 \leq 0. \label{ws} 
\end{align}


Since the UAV is assumed to fly along a Manhattan trajectory, the weight on each edge contains the same completion time-related component $T_r$. Despite this invariance, the term $T_r$ is indispensable for exploring the trade-offs between completion time and the other two objectives. The outage-related term $O_r$ on an edge is determined by the outage status of both grid points (vertices) the edge connects. Note that the outage status of an unmeasured grid point is determined by the estimated SINR based on Kriging interpolation. The measurement-related term $M_r$ demonstrates the directionality of the graph, which is determined solely by the measurement status of the end vertex, and therefore the UAV is encouraged to fly along the directions of the edges toward unmeasured gird points and avoid measured ones. Specifically, $|V(D_s)|=I\cdot J\cdot K$, $|E(D_s)|=2((I-1)\cdot J\cdot K+I\cdot (J-1)\cdot K+I\cdot J\cdot (K-1))$. Each $v\in V(D_s)$ is a balanced vertex and $D_s$ is a balanced Eulerian digraph. Define an indicator function for each $(v_a,v_b) \in E(D_s)$ as 
\begin{equation}
x_{ab} = 
\begin{cases}
1, \quad &\text{$\boldsymbol{u}_r[i]=v_a,\boldsymbol{u}_r[i+1]=v_b\quad  \exists i\in\{1,2,..,N-1\}$}  \\ 
0.  
\end{cases}      
\end{equation} to signal if the UAV traverses the edge $(v_a,v_b)$ during its flight. Therefore, (P1) can be reformulated as follows.

\begin{align}
\textup{(P2):}\quad  \notag 
\min_{\boldsymbol{u}_r[n],1\leq n \leq N} &\quad \sum_{(v_a,v_b) \in {E(D_s)}} w_{ab}^s \cdot x_{ab} \\
\text{s.t.} \quad 
& \{\boldsymbol{u}_r[1], \boldsymbol{u}_r[2],...\boldsymbol{u}_r[N]\}\rightarrow V(D_s),  \label{arrow} \\
&\sum_{v_b\in V(D_s)\backslash\{v_s\}} x_{sb} = 1,  \quad  v_s\in V(D_s), \label{start} \\
&\sum_{v_a\in V(D_s)\backslash\{v_e\}} x_{ae} = 1,  \quad  v_e\in V(D_s), \label{end}\\
&\sum_{v_i \in V(D_s)\backslash\{v_k\}} x_{ik} = \sum_{v_j \in V(D_s)\backslash\{v_k\}} x_{kj} \quad  \notag \\
&\forall v_k \in V(D_s) \setminus \{v_s, v_e\},   \label{propagation} \\
&u_r[1]=v_s, \\
&u_r[N]=v_e. 
\end{align}

The objective to be minimized is the sum of weights on the UAV's trajectory; ``$\rightarrow$'' in \eqref{arrow} means that the map from $\{\boldsymbol{u}[1], \boldsymbol{u}[2],...\boldsymbol{u}[N]\}$ to $V(D_s)$ is injective; $x_{ab}$ is a binary variable used to indicate if the edge $(v_a,v_b)$, as in this order, is part of the UAV's trajectory; \eqref{start} and \eqref{end} define the starting point and the end point; \eqref{propagation} ensures that the trajectory is a path. Note that the left-hand side and the right-hand side of \eqref{propagation} equal 1 only when $v_k$ corresponds to one of the $\boldsymbol{u}_r[n], 1\leq n\leq N$. Due to the nonpositivity of $\mu_2$, the weight values on some edges might be negative. Therefore, traditional shortest-path algorithms like Dijkstra, which can only deal with positive weight values, cannot be directly applied. The problem is solved via the Floyd algorithm, whose time complexity is $\mathcal{O}(|V(D_s)|^3)$.

\subsection{The TSP-based Solutions}

The traditional TSP is formulated as follows\cite{TSP}. A traveling salesman wishes to visit $N \geq 3$ cities $\mathcal{S}=\{v_1,v_2,...,v_N\}$, and return to his starting city $v_s\in \mathcal{S}$ while covering the least possible total distance. Each city can only be visited once, and therefore the route of the salesman is a ``Hamiltonian cycle'' of $K_N(\mathcal{S})$, where the weight on each edge is the distance between the two cities the edge connects. Open TSP is a variant of the TSP in which the starting city $v_s\in \mathcal{S}$ and the end city $v_e\in \mathcal{S}$ are different, and the salesman needs to find the shortest ``path'' from $v_s$ to $v_e$ that traverses all cities in $\mathcal{S}$. 

As pointed out in Section III, (P1) consists of two major aspects, namely UAV navigation, and CKM completion. Consider a scenario in which the operator of the UAV assigns relatively higher priority on CKM completion over UAV navigation. Then, the most ``radical'' strategy is to design a trajectory that traverses all unmeasured grid points in a single round of flight while minimizing the completion time and outage duration. Specifically, the unmeasured grid points can be considered as cities and the UAV can be regarded as the salesman trying to traverse them. Due to different starting and end points, the trajectory design process fits into the Open TSP. 

However, incorporating all unmeasured grid points into the set $\mathcal{S}$ is infeasible for two reasons. First, Open TSP is an NP-hard problem whose complexity grows exponentially with the scale of the problem, i.e., $|\mathcal{S}|$. Therefore, an overly large $|\mathcal{S}|$ would cause the problem to be practically unsolvable. Second, even if the trajectory is successfully determined, it might be kinematically infeasible due to the sharp acceleration and deceleration needed to traverse densely-distributed unmeasured grid points.

A milder strategy is to select a subset of unmeasured grid points with the highest ``measurement value'' to constitute the set $\mathcal{S}$ in each round, and iteratively update the global CKM. The ``measurement value'' of each unmeasured grid point can be determined based on the variance of estimation errors predicted by Kriging. Specifically, Denote the set of all unmeasured grid points in the $r^{th}$ round of flight as $\mathcal{U}_r$, and the set of all measured grid points as $\mathcal{K}_r$. Suppose that the UAV measures a set of grid points $\mathcal{S}_r\subseteq \mathcal{U}_r$, then $\mathcal{U}_{r+1}=\mathcal{U}_r\setminus \mathcal{S}_r$, $\mathcal{K}_{r+1}=\mathcal{K}_r\cup \mathcal{S}_r$. Specifically, if we use $\mathcal{K}_r\cup \mathcal{S}_r$ to estimate the SINR in $\mathcal{U}_r\setminus \mathcal{S}_r$, then the variance of estimation error $\epsilon_{\boldsymbol{u}}$ in an unmeasured grid point $\boldsymbol{u}\in \mathcal{U}_r\setminus \mathcal{S}_r$ is $\sigma_{\boldsymbol{u}}^2(\mathcal{K}_r\cup \mathcal{S}_r) = \boldsymbol{r}_0^T \boldsymbol{R}^{-1} \boldsymbol{r}_0$ \eqref{Matrix}. The selection of $\mathcal{S}_r$ can be formulated as the following optimization problem 

\begin{align}
\textup{(P3):}\quad 
\min_{\mathcal{S}_r} &\sum_{\boldsymbol{u} \in \mathcal{U}_r \setminus \mathcal{S}_r} \sigma_{\boldsymbol{u}}^2(\mathcal{K}_r \cup \mathcal{S}_r) \\
\text{s.t.} &|\mathcal{S}_r| = N, 
\\ &\mathcal{S}_r \subseteq \mathcal{U}_r.
\end{align}

In order to minimize the global MSE in (P1), the optimization objective is defined as the sum of the estimation variance in all unmeasured grid points $\mathcal{U}_r \setminus \mathcal{S}_r$ based on the SINR in $\mathcal{K}_r \cup \mathcal{S}_r$. Since $\text{(P3)}$ is an NP-hard combinatorial mathematics problem, there is no exact polynomial algorithm to solve it. To tackle this problem, an approximate greedy algorithm is proposed and summarized in Algorithm 1.

\begin{breakablealgorithm}
\caption{Greedy algorithm for solving (P3)}
\begin{algorithmic}[1] 
\State \textbf{Initialize:} Obtain the sets $\mathcal{K}_r$ and $\mathcal{U}_r$ in the $r^{th}$ flight. $\mathcal{S}_r \gets \emptyset$.
\For{$n \gets 1$ to $N$}  
\For{each $s_i\in \mathcal{U}_r$}  
\State Calculate $\mathcal{J}_i=\sum_{\boldsymbol{u} \in \mathcal{U} \setminus \mathcal{S}_r\setminus \{s_i\}} \sigma_{\boldsymbol{u}}^2(\mathcal{K}_r \cup \mathcal{S}_r \cup \{s_i\})$.
\EndFor 
\State $m \leftarrow \arg\min_{i \in \{1,2,...|\mathcal{U}_r\setminus\mathcal{S}_r|\}} \mathcal{J}_i$.
\State $\mathcal{S}_r\leftarrow \mathcal{S}_r \cup\{s_m\}$.
\State $\mathcal{U}_r\leftarrow \mathcal{U}_r \setminus\{s_m\}$.
\State $n\leftarrow n+1$.
\EndFor 
\State \Return $\mathcal{S}_r$.
\end{algorithmic}
\end{breakablealgorithm}

Since before the $r^{th}$ round of flight, we have $|\mathcal{U}_r|=U$, $|\mathcal{S}_r|=N$, $|\mathcal{K}_r|=K$. The number of iterations of the outer loop is $N$, the number of iterations of the middle loop is $U$. The number of candidate grid points to evaluate in the $i^{th}$ outer loop is $U-(i-1), 1 \leq i\leq N$. Given that $N$ is usually much smaller than $U$, the number of iterations of the middle loop is approximately $N\cdot U$. In terms of the inner core calculation of solving the Kriging covariance matrix \eqref{Matrix}, the time complexity is $\mathcal{O}((K+N)^3)$. Therefore, the overall time complexity for solving (P3) is $\mathcal{O}(N\cdot U\cdot (K+N)^3)$.

In order for the TSP-based navigation strategy to work, one major factor to consider is the spatial distribution pattern of the high-quality grid points as the solution to (P3). Due to the second-order stationarity assumption of Kriging, reliable predictions, i.e., low variances of estimation error, are only feasible for unmeasured grid points located in the proximity of measured grid points. Specifically, when the distance between them exceeds the range $a$ in \eqref{semivariogram}, the spatial correlation becomes negligible, and large variance of estimation error occurs. Driven by the optimization objective in (P3), the greedy algorithm would preferentially constitute $\mathcal{S}_r$ using grid points in regions with the highest Kriging variance. If such regions are distributed sparsely, the resulting $\mathcal{S}_r$ would be too scattered to serve as the UAV's waypoints. 

One way to address the above issue is to reduce $\mathcal{U}_r$ in (P3) to a subset of all unmeasured grid points. Specifically, the distance from potential high-quality grid points to the line connecting the starting and end points should not exceed a threshold so that the UAV would not severely deviate from its flight mission. Such a strategy prioritizes the estimation accuracy in grid points along the UAV's major flight routes, defined by the locations of the starting and end points.

Therefore, for the TSP-based navigation strategy, the tradeoff between $M_r$ and $T_r, O_r$ could be reflected by adjusting the cardinality of $\mathcal{S}_r$, i.e., $N$, and the allowable deviation distance. The weight mapping function can focus on the expense in terms of $T_r$ and $O_r$. Suppose that $\mathcal{S}_r=\{v_1,v_2,...v_N\}$, where $v_a=\boldsymbol{u}^D(i_a,j_a,k_a)$, then the distance from $\forall \boldsymbol{u}^D(i,j,k)$ to the line segment $v_a-v_b$ is $dist^{a,b}_{i,j,k}=\frac {\| (v_a-\boldsymbol{u}^D(i,j,k))\times(v_a-v_b)\|}{\|v_a-v_b\|}$. Similar to the definition in Section III, the approximated length of intersection between $v_a-v_b$ and the grid point ${u_{i,j,k}^D}$ is $L^{a,b}_{i,j,k}=2\sqrt{R^2-{(dist^{a,b}_{i,j,k})}^2}$. Let $G_t=(V(G_t),E(G_t),\boldsymbol{w}_t)$ be an undirected graph, where the subscript $t$ refers to the ``TSP-based''. Then, we have 

\begin{align}
&V(G_t)=\mathcal{S}_r, \\
&E(G_t)=E(K_N(\mathcal{S}_r)),
\\
&\boldsymbol{w}_t((v_a,v_b))= w_{a,b}^t=\notag
\\&\underbrace{\|v_a-v_b\|}_{T_r}+\underbrace{\beta\cdot(\sum_{i = 1}^{I} \sum_{j = 1}^{J} \sum_{k = 1}^{K}  I_T(dist^{a,b}_{i,j,k}) \cdot I_O(u_{i,j,k}^D) \cdot L^{a,b}_{i,j,k})}_{O_r},\beta\geq 0. \label{wt}
\end{align}

The Open TSP is thus formulated as follows.

\begin{align}
\textup{(P4):}\quad \notag  
\min_{\boldsymbol{u}_r[n],1\leq n \leq N} &\quad \sum_{(v_a,v_b) \in E(G_t)} w_{ab}^{t} \cdot x_{ab} \\
\text{s.t.} \quad 
& \{\boldsymbol{u}_r[1], \boldsymbol{u}_r[2],...\boldsymbol{u}_r[N]\}\leftrightarrow \mathcal{S}_r, \label{first}   \\
&\sum_{v_b\in V(G_t)\backslash\{v_s\}} x_{sb} = 1,  \quad  v_s\in \mathcal{S}_r, \label{start2} \\
&\sum_{v_a\in V(G_t)\backslash\{v_e\}} x_{ae} = 1,  \quad  v_e\in \mathcal{S}_r, \label{end2} \\
&\sum_{v_a\in V(G_t)\backslash\{v_s,v_e\}} x_{ab} = 1,  \quad  \forall v_b \in \mathcal{S}_r \backslash\{v_s,v_e\}, \\
&\sum_{v_b\in V(G_t)\backslash\{v_s,v_e\}} x_{ab} = 1,  \quad  \forall v_a \in \mathcal{S}_r \backslash\{v_s,v_e\}, \\
&\sum_{v_a \in \mathcal{S}_r} \sum_{v_b \in \mathcal{S}_r} x_{ab} = N - 1,   \label{loop1} \\
&\sum_{v_a \in S} \sum_{v_b \in S} x_{ab} \leq |S| - 1, \quad \notag \\ 
&\forall S \subset \mathcal{S}_r, \, 2 \leq |S| \leq N-1, \label{loop2} \\
&\boldsymbol{u}_r[1]=v_s, \\
&\boldsymbol{u}_r[N]=v_e. \label{last}
\end{align}

``$\leftrightarrow$'' in \eqref{first} means that the map from $\{\boldsymbol{u}_r[1], \boldsymbol{u}_r[2],...\boldsymbol{u}_r[N]\}$ to $\mathcal{S}_r$ is invertible so that each grid point in $\mathcal{S}_r$ is visited only once; \eqref{start2} and \eqref{end2} define the starting point and the end point; \eqref{loop1} and \eqref{loop2} jointly prevent the generation of short cycles.

One of the standard solutions to an Open TSP is to transform the problem into an equivalent TSP. Denote the weight between $v_s$ and $v_e$ as $w_{se}$. The approximated equivalency is realized by manually revising the value of $w_{se}$ to make sure that the edge $(v_s,v_e)$ is in the solution set of the equivalent TSP problem. The edge $(v_s,v_e)$ is deleted in the end to break the cycle and obtain a feasible path. For the sake of strictness, suppose that the smallest weight on $E(G_t)\setminus\{(v_s,v_e)\}$ is $s$, and the largest weight on $E(G_t)$ is $l$. As long as $w_{se} < (N-1)s-(N-2)l$, the edge $(v_s,v_e)$ is bound to be in the solution of the equivalent TSP. The equivalent TSP can be solved by various approximate polynomial algorithms, such as the classic Christofides algorithm\cite{TSP}, whose time-complexity is $\mathcal{O}(N^3)$. In addition, heuristic algorithms such as simulated annealing (SA)\cite{SA}, particle swarm optimization (PSO)\cite{Particle}, and their variants can be exploited to explore the lower bound of the TSP-based solution.

\section{Simulation Results}

In this section, simulation results are provided to verify the performance of the proposed algorithms. Given the importance of the SPP in trajectory planning\cite{zhan2022energy}\cite{Zhang}, we first evaluate the performance of the SPP on the short and long time scales. Then, we present the Pareto boundaries of the two algorithms to explore their major characteristics, and discuss potential ways to combine them. The urban environment consisting of high-rise buildings with different locations and heights is generated based on one realization of the statistical model suggested by International Telecommunication Union (ITU) \cite{ITU}. The link state (LoS/NLoS) from BS $m$ to each grid point $u_{i,j,k}^D$ can be determined by checking if there is any building blocking the line connecting $m$ and $u_{i,j,k}^D$. The simulation parameters are summarized in Table II.

\begin{table}[htbp]
\centering
\caption{Simulation parameters}
\begin{tabularx}{\linewidth}{p{0.75\linewidth}  
    X  
  }
\hline
\textbf{Simulation Parameter} & \textbf{Value} \\ \hline
The spatial span along the x-axis $|x_U-x_L|$. & 2000 m   \\ \hline
The spatial span along the y-axis $|y_U-y_L|$. & 2000 m   \\ \hline
The spatial span along the z-axis $|z_U-z_L|$. & 100 m   \\ \hline
The discretization granularity$\Delta_D$. & 10 m  \\ \hline
The speed of the UAV $V_{max}$. & 10 m/s \\ \hline
The number of ground BSs $M$. & 7 \\ \hline
The height of each BS $H_m$. & 25 m \\ \hline
The noise power at the receiver $\sigma_n^2$. & -110 dBm \\ \hline
The number of array element $N_0$ in each BS. & 8 \\ \hline
The down-tilt angle $\theta_D$ of each BS. & $10^\circ$ \\ \hline
The half-power bandwidth $\Theta_{3dB}$ of each array element. & $65^\circ$ \\ \hline
The threshold for antenna nulls $G_0$. & 30 dB \\ \hline
The communication bandwidth $B$. & 1 MHz \\ \hline
The carrier frequency $f_c$. & 3 GHz \\ \hline
\end{tabularx}
\end{table}

Fig. 4(a) shows the ground truth CKM and Fig. 4(b) presents the partial CKM with $50\%$ data missing. Note that the CKM predicts the SINR in each grid point in a statistical manner. When the UAV traverses a grid point that is considered not in outage based on the CKM, it might still experience outage due to the randomness of small-scale fading, which is faithfully recorded in the simulation.

\begin{figure}[H]  
    \centering  
    
    \begin{subfigure}[b]{0.45\textwidth}
        \centering
        \includegraphics[width=8cm,height=6cm]{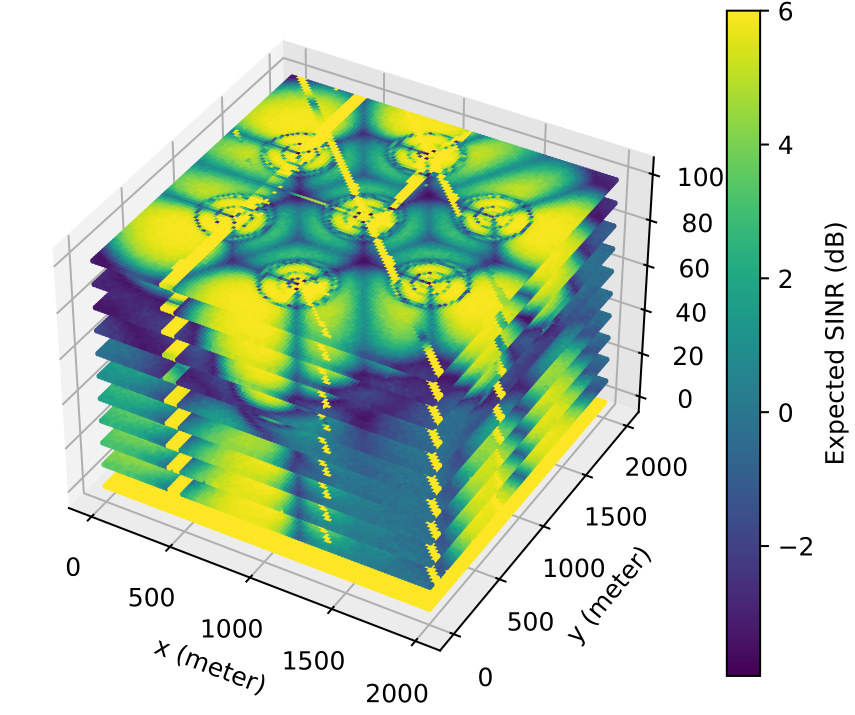}
        \subcaption{The ground truth CKM.}  
    \end{subfigure}
    
    \vspace{5pt}  

    \begin{subfigure}[b]{0.45\textwidth}
        \centering
        \includegraphics[width=8cm,height=6cm]{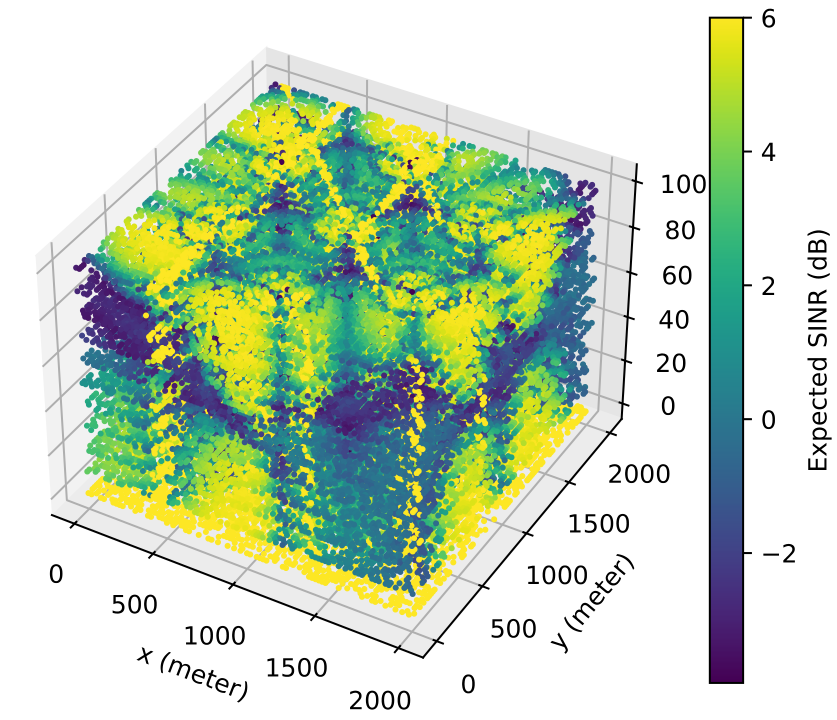}
        \subcaption{The partial CKM.}
    \end{subfigure}
    
    \caption{Visualization of CKM.}  
    \label{fig:main}  
\end{figure}

\subsection{Evaluation of the SPP}

The structure of the directed graph $D_s$ allows the UAV to finely adjust its trajectory to avoid coverage holes and traverse previously unmeasured regions. For the short time scale performance, Fig. 5 shows the SPP-based trajectories in a single round of flight with different $\mu_1$. For the sake of clarity, the $\mu_2$ of these trajectories is set to zero. As can be observed, when $\mu_1=0$, the UAV's trajectory is the standard Manhattan route that minimizes the completion time $T_r$ while ignoring the $O_r$ and $M_r$ along the way. As $\mu_1$ increases, the trajectory becomes increasingly tortuous to avoid grid points in outage.

\begin{figure}[H]	
\centering	
\includegraphics[width=9cm,height=6.5cm]{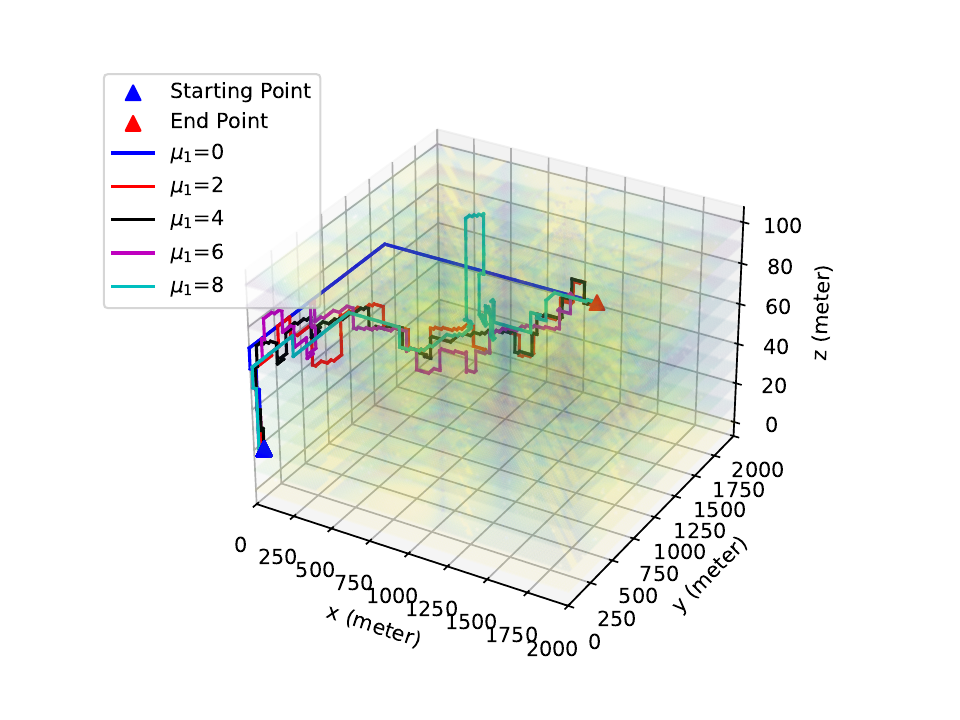}	
\caption{Trajectories based on the SPP.}
\end{figure}

\begin{figure}[H]	
\centering	
\includegraphics[width=8cm,height=6cm]{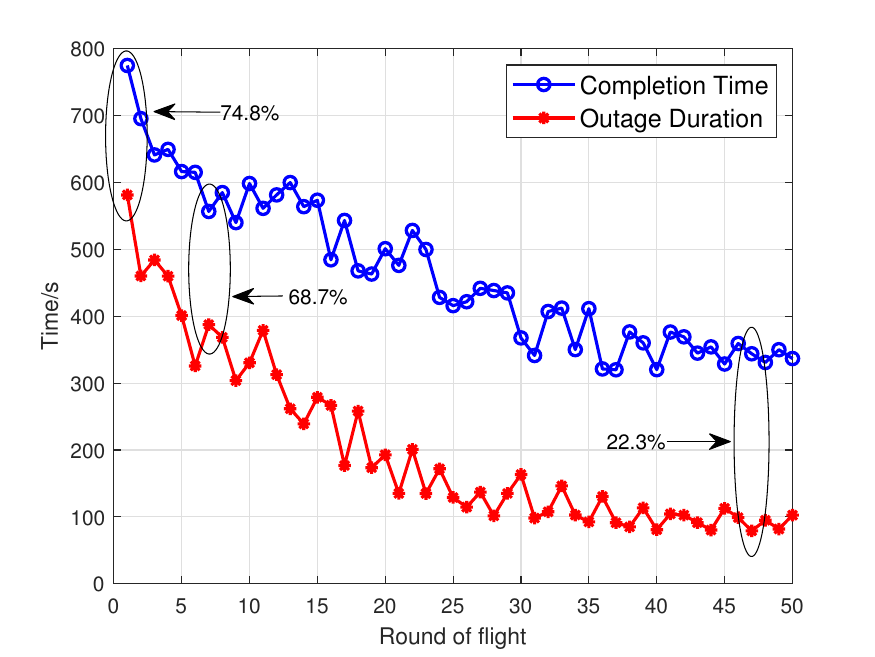}	
\caption{Performance evaluation of the SPP-based strategy,$\gamma_{th}=0 \text{dB}, \mu_1=8, \mu_2=-3$.}
\end{figure}

In terms of the long time scale performance, Fig. 6 shows the completion time and outage duration versus the round of flight based on the SPP. The starting and end points are identical for all rounds. It is observed that both $T_r$ and $O_r$ decrease as the flight round increases. The mechanism for this change is that as more previously unknown grid points become measured, more grid points having good communication quality are revealed, reducing the number of unnecessary detours and the total outage distance. Moreover, the proportion of the outage duration to the total completion time also decreases as the round of flight increases. Specifically, in early rounds of flight, the proportion can be as high as $74.8\%$, while after around 30 rounds, the percentage converges at approximately $22\%$. The result demonstrates that resorting solely to the original partial CKM only yields an unsatisfactory trajectory, in the sense that both $T_r$ and $O_r$ are relatively high. With a moderate allowance of deviation ($\mu_2=-3$), the newly measured CKM information can rapidly expand the Pareto boundary of the problem. It is also worth mentioning that the two curves in Fig. 6 do not decrease monotonically. The reason for this phenomenon is that after the UAV's $r^{th}$ round of flight, the measurement indicators for the newly measured grid points during the $r^{th}$ round turn into 1, and the term $M_r$ in \eqref{ws} correspondingly increases. Driven by the smaller $M_r$ in unmeasured grid points, the completion time and outage duration in the $r+1^{th}$ round might be larger than that in the $r^{th}$ round due to the extra detours to traverse unmeasured grid points that are potentially in outage.

\subsection{Pareto Boundary Analysis}

In this subsection, the Pareto boundaries are analyzed to explore the trade-offs among the three short-term optimization objectives. Specifically, three pairwise combinations of the three optimization objectives are constructed to form tradeoff curves. Each data point on these curves corresponds to the same flight mission, in the sense that the starting/end points, and the outage threshold are identical. For the SPP-based strategy, the trade-off curves are generated by adjusting $\mu_1$ and $\mu_2$ in \eqref{ws}. For the TSP-based strategy, they are generated by adjusting $|\mathcal{S}_r|$ and $\beta$ in \eqref{wt}.

\begin{figure}[H]	
\centering	
\includegraphics[width=9cm,height=7cm]{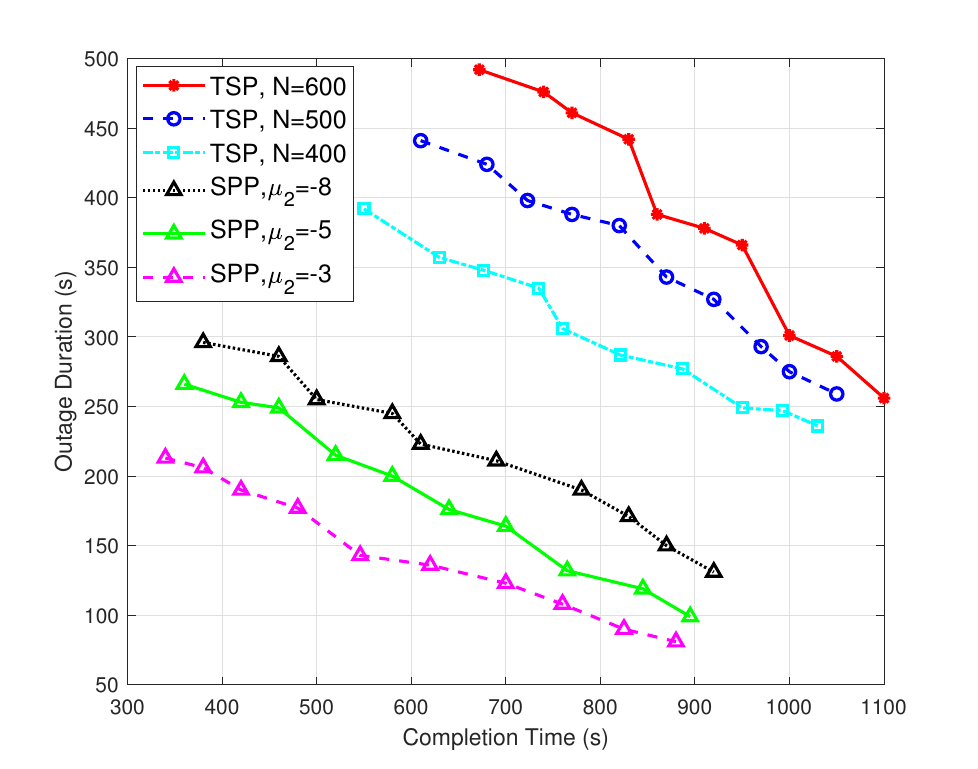}	
\caption{Trade-off between completion time and outage duration.}
\end{figure}

Fig. 7 shows the trade-off between $T_r$ and $O_r$. For the SPP, curves of different colors correspond to different values of $\mu_2$ to characterize the influence of $M_r$, and the data points on each curve are generated by adjusting $\mu_1$ in \eqref{ws}. For the TSP, the data points on the same curve correspond to the same vertex set $\mathcal{S}_r$, and the tradeoff is explored by adjusting $\beta$ in \eqref{wt}. For both algorithms, an increase in $T_r$ leads to a decrease in $O_r$. For the SPP, as $\mu_1$ increases, the UAV is more inclined to avoid grid points that are in outage, leading to more detours and therefore longer completion time. Similarly for the TSP, as $\beta$ increases, the UAV is encouraged to adjust its traversal sequence of $\mathcal{S}_r$ to minimize $O_r$ at the expense of prolonged $T_r$.

It is noteworthy that the SPP-based curves exhibit a nearly linear trend of decrease for all $\mu_2$. The reason for this gentle change can be due to the Manhattan constraint for the UAV's trajectory. Specifically, the definition of $E(D_s)$ fundamentally restricts the UAV's trajectory in each step to a local spatial neighborhood, precluding the UAV from implementing large-scale detours to reduce $O_r$ in a global manner. Consequently, any incremental decrease in $O_r$ can only be achieved through small, stepwise extensions of the UAV's local path, leading to a gradual, rather than abrupt, increase in $T_r$. For the TSP, on the other hand, the spherical approximation unlocks the UAV's mobility, and therefore the adjustment in $\beta$ could lead to sharp turning points in the TSP-based curves, corresponding to major structural reconstruction of the UAV's trajectory to avoid potential coverage holes.

\begin{figure}[H]	
\centering	
\includegraphics[width=9cm,height=7cm]{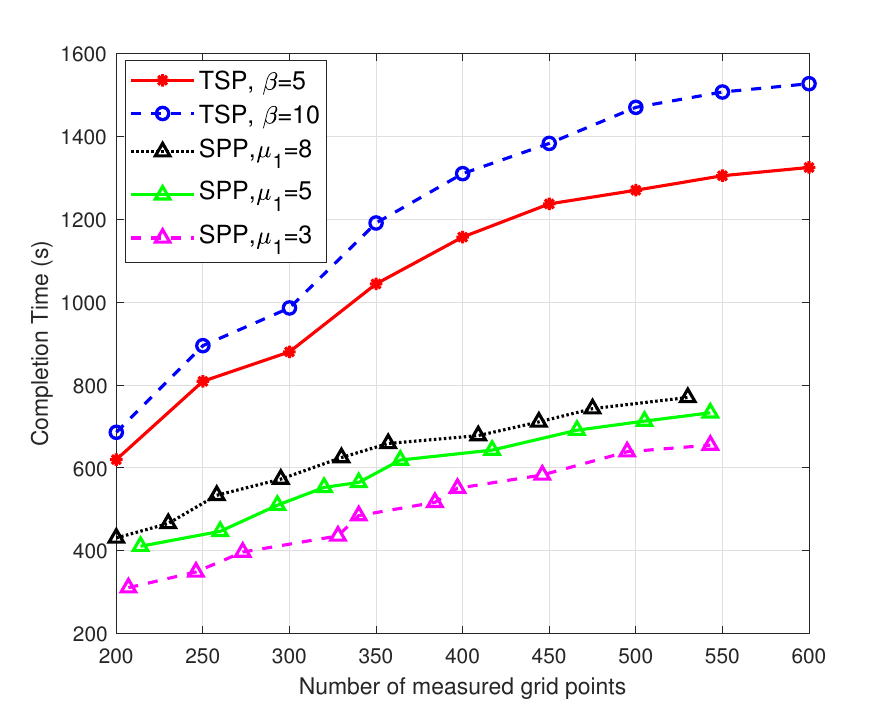}	
\caption{Trade-off between completion time and number of measured grid points.}
\end{figure}

Fig. 8 shows the trade-off between $T_r$ and $M_r$. For the SPP, as $|\mu_2|$ increases, the UAV is encouraged to traverse more previously unmeasured grid points, resulting in larger $T_r$. Due to the same Manhattan constraint, the SPP-based curves exhibit a linear trend of increase. On the contrary, the TSP-based curves exhibit clear piecewise characteristic, where the slope of the curves decreases as $N$ increases. The reason for this characteristic could be explained as follows. When $N$ is small, the UAV is only allowed to traverse a limited number of unmeasured grid points. As pointed out in Section IV, Kriging, in this regime, is inclined to identify high-quality unmeasured grid points in isolated, sparsely distributed high-variance regions to achieve a significant global information gain. The newly added points in $\mathcal{S}_r$ tend to fill in previously uncovered high-variance regions located far away from covered regions, and therefore the total traveling distance grows rapidly. As $N$ increases, the semivariogram model becomes robust enough to capture the spatial correlation over a larger spatial extent, allowing Kriging to accurately estimate the SINR across a broader area. The high-quality grid points are no longer confined to local clusters, and the newly added grid points fill in the middle regions between the original high-variance regions. Therefore, $\mathcal{S}_r$ is more uniformly distributed across the 3D aerial space, and the UAV only needs to take small detours to traverse the newly added grid points. Consequently, the TSP-based tradeoff curves exhibit a gentle slope, which is similar to that of the SPP, when $N$ is relatively large. In addition, Fig. 9 shows the trade-off between $O_r$ and $M_r$. The SPP-based curve is presented by adjusting both $\mu_1$ and $\mu_2$. The TSP-based curves exhibit the same piecewise slopes as in Fig. 8. Due to the fact that some of the unmeasured grid points are also in outage, the percentage of $O_r$ to $T_r$ increases with $N$, which applies to both algorithms.

\begin{figure}[H]	
\centering	
\includegraphics[width=9cm,height=7cm]{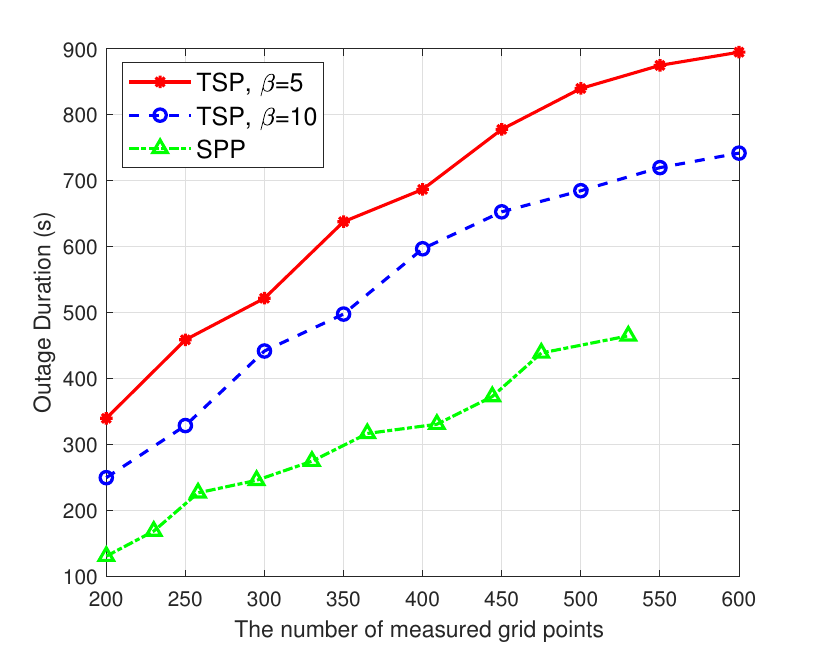}	
\caption{Trade-off between outage duration and number of measured grid points.}
\end{figure}

\begin{figure}[H]	
\centering	
\includegraphics[width=9cm,height=7cm]{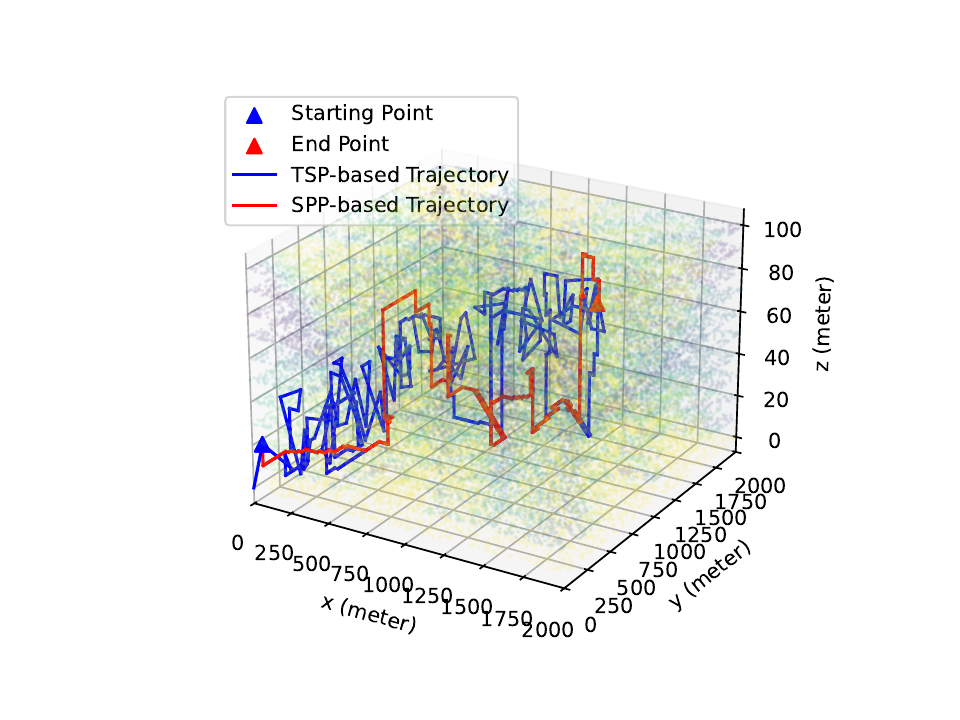}	
\caption{Visual comparison between the TSP-based and SPP-based trajectories.}
\end{figure}

Fig. 10 illustrates the trajectories of the two strategies with the same flight mission. The TSP-based trajectory has to make more sharp turns in order to traverse the sparsely distributed high quality grid points, while the SPP-based trajectory appears to be more regular due to the Manhattan constraint. Note that for a low quality unmeasured grid point $\boldsymbol{u} \in \mathcal{U}_r \setminus \mathcal{S}_r$, it could still be traversed and measured by the UAV when it intersects with one of the $N-1$ line segments of the UAV's trajectory, suggesting that the total number of newly measured grid points during one round of flight based on the TSP could be larger than the designated number of waypoints $N$.

Lastly, regarding CKM completion, the UAV is required to execute $R=50$ rounds of flight, and the available CKM prior to each round of flight is based on the original CKM and the data measured in previous rounds. For both the SPP and TSP, after each round of flight, the SINR in unmeasured grid points is estimated based on the SINR in measured grids and Kriging interpolation. The starting points of the 50 rounds of flight are randomly generated across the 3D aerial space, while the end points are identical so as to evenly cover the unmeasured regions of the CKM. 

\begin{figure}[H]	
\centering	
\includegraphics[width=9cm,height=7cm]{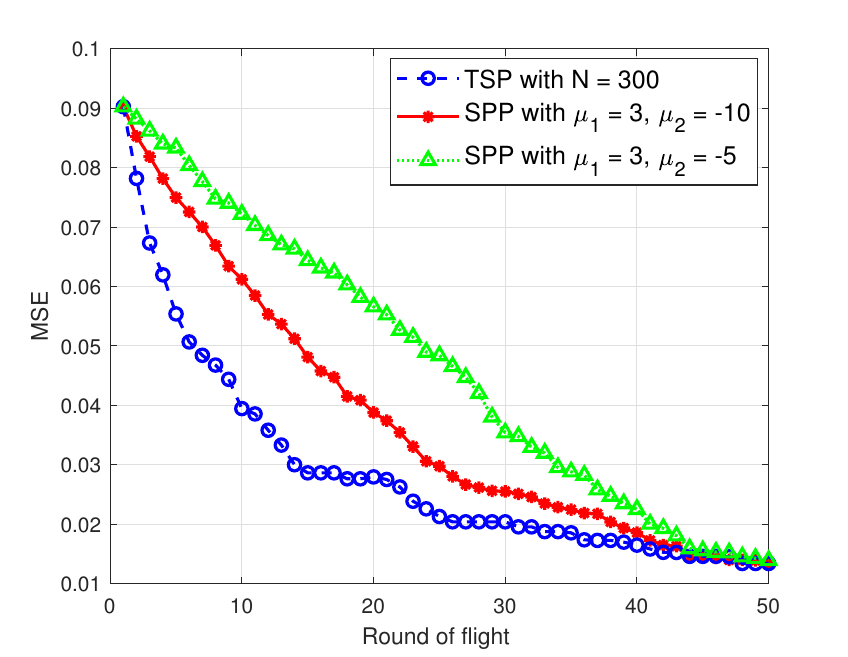}	
\caption{The MSE versus the round of flight for the SPP and TSP.}
\end{figure}

Fig. 11 shows the MSE versus the round of flight for the SPP and TSP. It is observed that the MSE of the TSP-based strategy decreases rapidly and converges around the $35^{th}$ round. The curves of the SPP-based strategy tend to be more linear, suggesting that the grid points they measure only provide a limited information gain. In addition, the SPP with larger $\mu_2$ assigned in \eqref{ws} converges more quickly. Moreover, the MSE of all three curves converged to 0.013. The reason for this limit can be due to the second-order stationarity assumption of Kriging, and better interpolation accuracy demands more advanced interpolation methods or artificial-intelligence (AI)-based strategies.

As can be seen from the above results, the SPP-based strategy is superior to the TSP-based method in terms of the completion time and outage duration. On the other hand, the obtained high-quality grid points based on (P3) allows the TSP-based strategies to rapidly complete the partial CKM. Therefore, the wide range between the upper and lower Pareto boundaries respectively set for the three optimization objectives can be explored by properly combining the two strategies.

The two strategies are applicable to different scenarios. Specifically, one particular sub-aerial space might be densely populated with high quality grid points and therefore it makes sense to use the TSP-based strategies to obtain a significant information gain. On the other hand, when the UAV's original flight mission subjects to stringent constraints regarding completion time or outage duration, the SPP-based strategy can be employed with a moderate $\mu_2$ to incidentally traverse some unmeasured grid points without compromising the original mission. Therefore, one simple way to combine the two strategies is to concatenate them in chronological order according to the structure of the partial CKM and the restrictions of flight missions.

More advanced combination can be realized through the revision of the UAV's flying pattern. In (P4), the UAV is assumed to fly in straight lines to connect the high-quality grid points in order, which limits the TSP-based strategies' ability to optimize the outage duration since the UAV is not allowed to take any detour when flying from $\boldsymbol{u}_r[i]$ to $\boldsymbol{u}_r[i+1]$. Therefore, after the acquisition of the set $\mathcal{S}_r$, one can change the flying pattern of the UAV from the straight line to the Manhattan trajectory, and subsequently employ the SPP to determine the shortest paths between any $v_a,v_b\in\mathcal{S}_r$. Such a sophisticated strategy applies to the case where the high-quality grid points are sparsely distributed, rendering the TSP-based broken-line trajectory unreasonably simple.

\section{Conclusion}

In this paper, a multi-objective optimization problem is formulated to combine UAV navigation and CKM completion into a common framework. The effectiveness of UAV navigation is characterized by the short time scale metrics, namely the completion time, outage duration, and number of measured grid points in a single round of flight, while the performance regarding CKM completion is quantized via the global MSE based on Kriging interpolation and the data collected from multiple rounds of flight. Two navigation strategies based, respectively, on the SPP and the TSP in graph theory are comprehensively investigated. Simulation results suggest that the SPP-based strategy demonstrates dominant superiority in optimizing the completion time and outage duration, while the TSP-based strategy supported by the determination of high-quality grid points allows for the efficient completion of the partial CKM. The analysis of the Pareto boundaries reveals the inner trade-offs among the optimization objectives and also suggests possible ways to combine the two strategies to cater to different mission requirements in practice.

\bibliographystyle{IEEEtran}  
\bibliography{Partial_CKM_UAV}

\end{document}